\documentclass[a4paper,fleqn]{cas-dc}

\renewcommand{\arraystretch}{1.3}

\usepackage[numbers]{natbib}
\usepackage{amsmath,amssymb,amsfonts}
\usepackage{hyperref}
\usepackage{algorithmic}
\usepackage{float}
\usepackage{graphicx}
\usepackage{textcomp}
\usepackage{xcolor}
\usepackage{multirow}
\usepackage{graphicx}
\usepackage{adjustbox}
\usepackage{longtable}
\usepackage[normalem]{ulem}
\usepackage{listings}
\usepackage{xcolor}

\definecolor{codegreen}{rgb}{0,0.6,0}
\definecolor{codepurple}{rgb}{0.58,0,0.82}
\definecolor{backcolour}{rgb}{0.95,0.95,0.92}

\usepackage{microtype}

\begin{document}

\let\WriteBookmarks\relax
\def\floatpagepagefraction{1}
\def\textpagefraction{.001}

\shorttitle{A Domain-Driven Design Simulator for Business Logic-Rich Microservice Systems}
\shortauthors{Daniel Pereira and António Rito Silva}

\title[mode = title]{A Domain-Driven Design Simulator for Business Logic-Rich Microservice Systems}

\author[1]{Daniel da Palma Pereira}[orcid=0009-0004-8233-8941]
\ead{danielppereira2002@tecnico.ulisboa.pt}

\affiliation[1]{organization={INESC-ID, Instituto Superior Técnico, University of Lisbon},
  addressline={Rua Alves Redol 9},
  city={Lisboa},
  postcode={1000-029},
  country={Portugal}}

\author[1]{António {Rito Silva}}[orcid=0000-0001-9840-457X]
\ead{rito.silva@tecnico.ulisboa.pt}
\cormark[1]
\cortext[cor1]{Corresponding author}


\begin{abstract}
Developing business-logic-rich microservices requires navigating complex trade-offs between data consistency and distributed coordination. Although patterns like Sagas and Transactional Causal Consistency (TCC) provide mechanisms to manage distributed state, validating their behavior before production is challenging. Current architectural simulators prioritize network metrics over domain semantics, whereas industry frameworks demand full-scale infrastructure deployments, preventing early architectural experimentation.

To bridge this gap, we introduce a \textit{Domain-Driven Design} (DDD) microservice simulator that isolates core business logic from communication and transactional infrastructure. By modeling microservice systems around aggregates, the simulator allows developers to evaluate identical application code under varying consistency guarantees and network constraints. It features support for multiple transactional models (Sagas, TCC) and seamless transitions across diverse deployment topologies, ranging from centralized execution to fully distributed environments.

We validate the simulator through the implementation and rigorous concurrency testing of a complex, multi-aggregate microservice system. Through empirical benchmarks, we quantify the performance, coordination overhead, and resilience of different transactional models across localized and distributed execution environments. The findings confirm that the simulator minimizes developer effort while providing a powerful, deterministic environment for the shift-left validation and optimization of business logic implementation in microservice architectures.

\end{abstract}

\maketitle

\section{Introduction}
\label{sec:introduction}

The microservice architecture (MSA)~\cite{FowlerMicroservices15, Thones15} is widely used for designing a large number of software systems due to its advantages in scalability and support for small development teams. However, it is not without its own complexities~\cite{Neal21}, such as development complexity and the management of data consistency. The former relates to the complexity of the middleware (e.g., Kafka\footnote{\href{https://kafka.apache.org/}{https://kafka.apache.org/}}) used to implement microservice systems. The latter relates to distributed operations that must be coordinated to support system functionalities, which is a result of the CAP theorem~\cite{Fox99}, dictating a trade-off between consistency and availability. When considered together, these two challenges mean that data consistency issues are often only identified late in the development process. Consequently, a team might realize that the microservice architecture does not actually pay off only after an extensive development effort~\cite{Mendonca21}.

Therefore, it is worth applying the principles of "shift-left"~\cite{Miller17, Nader-Rezvani2019}, where verification starts as soon as possible in the development process, to the difficult aspects associated with the design and implementation of microservices, such as the complexity of handling distributed transactions~\cite{Santos20}. Developers must address the complexity of implementing business logic without traditional ACID atomic transactions. To achieve this, practitioners rely on alternative coordination models, such as Sagas~\cite{Garcia-Molina87, richardson19} (an application-level eventual consistency pattern), or follow more strict transactional models like Transactional Causal Consistency (TCC)~\cite{Wu20, akkoorath2016cure}. Following these models, developers must address problems resulting from a lack of isolation due to the existence of intermediate states, such as lost updates and dirty reads. Solutions like semantic locks, where transactions flag intermediate states for other transactions, require a case-by-case design that can be error-prone~\cite{richardson19}.

Domain-Driven Design (DDD)~\cite{Evans03} provides a robust conceptual framework for architecting microservice systems~\cite{Zhong24}. At its core, DDD involves decomposing the domain model into atomic units known as aggregates~\cite{Evans03}, which serve as the foundation for microservice construction. By identifying relevant aggregates through domain analysis, designers can delineate the transactional contexts of system functionalities. This, in turn, enables an analysis of how transactions are coordinated across the aggregates required to fulfill a specific business operation. However, such an analysis necessitates a complete specification of the aggregates' business logic. The impact of the lack of isolation is fundamentally determined by the specific operations being performed. For instance, a book's price might be temporarily outdated within a client's shopping cart, creating potential consistency issues if that intermediate state is visible to other processes. Identifying these scenarios is a critical component of microservice design, as they dictate both the system architecture and its underlying business operations. Such fundamental decisions should not be deferred until the final stages of development, they must be addressed early to reduce the impacts of change.

Several distributed system simulators have been proposed for the study of system qualities, such as performance and fault-tolerance. However, they largely overlook the complexities of microservice systems characterized by rich domain logic and DDD principles. Tools such as SimBlock~\cite{Simblock:Aoki:2019} and VWR~\cite{VWR:Chen:20} explicitly abstract away application-level logic to minimize computational overhead, focusing instead on service roles (like load balancers) or protocol-level message exchange. Performance-centric simulators like Terminus~\cite{terminus19}, $\mu$qSim~\cite{muqsim19}, and PerfSim~\cite{perfsim23} treat microservices as "black boxes" or "stages" defined by statistical time distributions and resource consumption. Consequently, these simulators do not account for the internal state transitions of DDD aggregates or the impact of transactional consistency models on business operations. By ignoring the semantic content of the business logic, current simulation approaches fail to support developers in experimenting with how domain-model boundaries and varying transactional guarantees influence the overall integrity and behavior of a microservice system.

Existing microservice frameworks, such as Eventuate Tram~\cite{eventuate_tram_about}, Axon Framework~\cite{axon_about}, Temporal~\cite{temporal_about}, and Conductor~\cite{conductor_about}, exhibit a significant architectural disconnect between operational coordination and domain-driven integrity. While these platforms provide robust environments for executing distributed business operations, they largely fail to treat domain-driven design as a first-class architectural concern. Furthermore, because these tools are primarily designed for production-level development, they do not provide a viable environment for a shift-left strategy, where architectural decisions can be simulated and validated before implementation.

To bridge the gap between theoretical modeling and real-world execution, this paper proposes a domain-driven design simulator for business-logic-rich microservice systems. The simulator facilitates the comprehensive modeling of complex services by explicitly applying DDD principles. It supports multiple transactional models for business operations and offers a controlled, centralized, and deterministic environment for shift-left verification. Furthermore, it can transition to a distributed context, closely approximating the behavior and execution of a production system.

Using this simulator, software architects can empirically assess the feasibility of the microservices architecture for the problem at hand, benchmarking different consistency strategies before committing to a final, expensive implementation.

The following research questions are addressed in this paper:

\begin{itemize}
    \item \textbf{RQ1}: How can a simulator facilitate Domain-Driven Design (DDD) for microservice systems, enabling developers to experiment with their models effectively?
    \item \textbf{RQ2}: How can a simulator enable experimentation with microservice business operations across one or several transactional models?
    \item \textbf{RQ3}: How can a simulator support diverse communication models while remaining decoupled from the underlying business logic?
    \item \textbf{RQ4}: What level of effort is required for a developer to model a microservice system and explore various supported configurations?
\end{itemize}

By addressing these questions, we provide a \textit{Microservice Simulator} that:

\begin{itemize}
    \item Directly incorporates Domain-Driven Design principles into the modeling process.
    \item Supports multiple transactional models through an extensible architecture that allows for easy integration of additional models.
    \item Accommodates various communication patterns (both centralized and distributed) and provides a framework for adding new ones.
    \item Decouples business logic from infrastructure, allowing developers to focus on domain logic while transactional and communication behaviors are handled via configuration.
    \item Supports the experimentation of the business logic behavior of a microservice system in a deterministic environment.
    \item Demonstrates expressive power through the implementation of a complex, logic-heavy system as a benchmark.
\end{itemize}

The remainder of this paper is organized as follows. Section~\ref{sec:background-related-work} defines the problem and reviews existing approaches. Section~\ref{sec:microservice-simulator} details the simulator's core infrastructural modules, while Section~\ref{sec:application_layer_quizzes} demonstrates how to instantiate a concrete microservice system within the framework. Supported deployment topologies are discussed in Section~\ref{sec:deployment_topologies}. Section~\ref{sec:evaluation} evaluates the simulator’s expressiveness regarding Domain-Driven Design (DDD) concepts, its application to a complex real-world system, and the required development effort. Furthermore, it assesses the simulator’s shift-left capabilities and provides a benchmark of various deployment topologies. Finally, Section~\ref{sec:conclusion} concludes the paper.

\section{Background and Related Work}
\label{sec:background-related-work}

This section provides the theoretical foundation for our work and reviews the current state of the art. We begin by examining the core principles of microservice architectures and the transactional models used to maintain data consistency in distributed environments. We then explore the tactical and strategic patterns of Domain-Driven Design (DDD), alongside the orchestration and communication mechanisms required to coordinate complex business logic across service boundaries. Finally, we evaluate existing frameworks and simulators, identifying the gaps that our proposed simulator aims to bridge.

\subsection{Microservice Architecture and Transactional Models}

The microservice architecture~\cite{FowlerMicroservices15, Thones15} is designed to address two primary challenges in software development: the effective implementation of agile methodologies within large organizations and the need for independent, demand-driven scalability. To address the first challenge, the application is decomposed into a suite of small services, each falling under the stewardship of a dedicated, cross-functional team. Regarding the second, these individual services can be scaled independently, allowing the system to allocate resources precisely where the current request load is heaviest without having to scale the entire application.

To achieve this, the application's domain model is decomposed into smaller, manageable segments. Consequently, the microservice architecture fundamentally decentralizes data management, ensuring that each service encapsulates its own state and operates as an independent unit~\cite{richardson19}. Therefore, traditional distributed transactions (e.g., two-phase commit or the XA standard) are widely considered anti-patterns due to the temporal coupling, synchronous communication, and performance bottlenecks caused by distributed locking. To achieve distributed data consistency while maintaining availability, two primary approaches have emerged:

\begin{itemize}
    \item \textbf{Sagas}: The Saga pattern is an application-level eventual consistency pattern initially conceptualized for long-lived transactions~\cite{Garcia-Molina87}. Instead of relying on distributed locks, a Saga coordinates a sequence of local ACID transactions using asynchronous messaging. If a local transaction fails, the Saga executes explicit compensating transactions to undo the preceding steps. Because Sagas inherently lack strict isolation (the `I' in ACID), intermediate states are exposed, leaving the system highly vulnerable to concurrency anomalies such as lost updates, dirty reads, and fuzzy (non-repeatable) reads~\cite{richardson19}. To mitigate these risks, developers must manually design conflict resolution countermeasures, such as semantic locks (e.g., flagging a record as \texttt{APPROVAL\_PENDING}), to manage intermediate states and prevent business rule violations.
    
    \item \textbf{Transactional Causal Consistency (TCC)}: Unlike Sagas, which serve as an architectural coordination pattern, TCC is a formal consistency model derived from distributed database theory that provides strict guarantees~\cite{akkoorath2016cure}. Specifically, TCC ensures two primary properties: transactions always observe a causally consistent snapshot of the system state, and updates to multiple objects enjoy atomic visibility~\cite{Wu20, Bravo21, lykhenko21, Toumlilt21}. By tracking causality, often via dependency matrices or vector clocks sized to the number of data centers, TCC automatically eliminates read-based concurrency anomalies by design. This offers a stronger, snapshot-based consistency model without sacrificing the high availability required by the CAP theorem. Furthermore, to automatically manage conflicting concurrent writes and ensure deterministic replica convergence, state-of-the-art TCC implementations pair this snapshot isolation with Conflict-Free Replicated Data Types (CRDTs)~\cite{Preguica2009, yu20}. While Sagas push the burden of anomaly resolution to the application developer, TCC leverages metadata and convergent data structures to resolve conflicts at the data store level, significantly reducing the complexity of the application's business logic.
\end{itemize}

\subsection{Domain-Driven Design}

Domain-Driven Design (DDD)~\cite{Evans03} provides a set of strategic and tactical patterns for managing software complexity. In the context of microservices, the most critical tactical pattern is the \textit{Aggregate}. An aggregate is a cluster of domain objects that can be treated as a single unit for data changes, enforcing business invariants within its boundaries. Every transaction in a DDD-compliant system should ideally modify only a single aggregate at a time, invoking one of its services. By modeling microservices around strict domain boundaries and deploying aggregates as isolated units, developers can logically partition a system. However, when a business requirement spans multiple aggregates (and therefore multiple microservices), developers must implement inter-aggregate coordination without violating the autonomy of the individual microservices.

The application of Domain-Driven Design in microservice systems design and implementation (DDD4M)~\cite{Zhong24} is currently characterized by a critical need for enhanced methodological support, specifically regarding domain knowledge acquisition and the practical design of domain models. According to an evidence-based investigation, while DDD4M is the go-to approach for projects with high functional complexity, practitioners frequently struggle to define clear microservice boundaries. Tactical patterns are currently applied with greater success than strategic ones, despite shortcomings in both. The aggregate tactical pattern is frequently omitted or, at best, inconsistently applied across the system. Regarding strategic patterns, the primary challenge lies in defining microservice boundaries, often not explicitly considering the internal business logic that tactical patterns are designed to capture.

\subsection{Microservice Orchestration}

Implementing comprehensive system features often requires \textit{Coordination}, which defines a long-running business transaction of the microservice system. This comprises an interaction between several aggregates to implement a microservice system functionality~\cite{Neal22}. 

This orchestration is managed through several structured concepts:
\begin{itemize}
    \item \textbf{Workflow}: It defines and controls the sequence of local transactions that constitute a microservice system's functionality, coordinating the invocations of various aggregate services~\cite{van2003workflow}.
    \item \textbf{Steps}: An abstract representation of a discrete unit of work within a larger microservice functionality. A workflow is composed of multiple steps, which encapsulate specific aggregate services.
    \item \textbf{Scheduling}: The mechanism responsible for the actual scheduling of workflow steps. It takes the raw steps and their dependency graphs to determine the correct topological execution order, ensuring prerequisites are met before subsequent steps begin.
    \item \textbf{Transactional Management}: It handles the management of transactional properties of the coordination of the interactions of several microservices in the context of microservice system functionality. \textit{Unit of Work} is a pattern that registers all the relevant changes that occurred during the workflow execution, and it is responsible for the final commit, or abort, of the functionality~\cite{fowler2012patterns}.
\end{itemize}

\subsection{Inter-Service Communication}

To maintain the autonomy of microservices while allowing them to collaborate, interactions between aggregates are managed through specific communication patterns. These interactions often form \textit{Upstream-Downstream} relationships\footnote{While DDD typically defines upstream-downstream relationships between \textit{Bounded Contexts}, which are groups of aggregates managed by a single team, the simulator shifts that focus. Since our emphasis is on coordinating business logic across a distributed environment where the aggregate serves as the primary unit of atomicity, we define these relationships as existing directly between individual aggregates.}, establishing development dependencies between the teams that manage them. The downstream team depends on, and is aware of, the upstream team, but not vice versa~\cite{evans2004domain}. 

Communication across these boundaries typically takes two forms:
\begin{itemize}
    \item \textbf{Events}: Used for upstream-to-downstream communication, domain events are emitted when an upstream aggregate changes. Downstream aggregates listen to these events to inform themselves of changes and react accordingly. Events act as triggers for the execution of functionalities in the downstream to react to the changes that occurred in the upstream aggregates, while preserving aggregate autonomy~\cite{evans2004domain}.
    \item \textbf{Commands}: While events handle reactive communication, commands are used for explicit downstream-to-upstream communication. Implementing the \textit{Command} pattern~\cite{gamma1994design}, they allow an orchestrator, through its steps, to explicitly dictate state changes or request actions from remote upstream services, effectively abstracting the underlying network transport protocols from the application's domain logic.
\end{itemize}

\subsection{Existing Frameworks}

Several industry frameworks facilitate the execution of distributed business workflows, including Eventuate Tram~\cite{eventuate_tram_about}, Axon Framework~\cite{axon_about}, and Temporal~\cite{temporal_about}. While these platforms provide highly robust environments for executing distributed business operations, they exhibit a significant architectural disconnect between operational coordination and domain-driven integrity. First, they lack native abstractions for DDD aggregates, requiring developers to manually bridge the semantic gap between their conceptual domain models and the framework's underlying infrastructure. Second, because these tools are primarily designed for production-level development, they do not provide a viable environment for a shift-left strategy. Mandating the deployment of full-scale infrastructure (e.g., databases, message brokers) prevents architects from utilizing the lightweight, deterministic environment necessary to simulate, validate, and debug complex coordination logic before implementation.

\subsection{Distributed System Simulators}


Several distributed system simulators have been proposed for the study of system qualities, such as performance and fault-tolerance. However, they largely overlook the complexities of microservice systems characterized by rich domain logic and DDD principles. Tools such as SimBlock~\cite{Simblock:Aoki:2019} and VWR~\cite{VWR:Chen:20} explicitly abstract away application-level logic to minimize computational overhead, focusing instead on service roles (e.g., load balancers) or protocol-level message exchange. Similarly, performance-centric simulators like Terminus~\cite{terminus19}, $\mu$qSim~\cite{muqsim19}, and PerfSim~\cite{perfsim23} treat microservices as "black boxes" defined solely by statistical time distributions and resource consumption. This infrastructure-centric focus is maintained by many microservice-specific simulators: OpenDC relies on statistical models for workload generation instead of executing business logic~\cite{ahsan2023opendc}, while telemetry tools like \texttt{microsim} generate synthetic traces via random latency mutations but omit aggregate state transitions~\cite{microsim}. Furthermore, simulators like MuSim focus on hardware utilization, modeling CPU load and memory bandwidth through synthetic benchmarks, while effectively abstracting away the underlying data state, transactions, and business rules~\cite{kanji2022scaling}.

Consequently, these simulators do not account for the internal state transitions of DDD aggregates or the impact of transactional consistency models on business operations. By ignoring the semantic content of the business logic, current simulation approaches fail to support developers in experimenting with how domain-model boundaries and varying transactional guarantees (e.g., Sagas versus TCC) influence the overall integrity of a microservice system. This gap necessitates a novel simulator that not only executes the actual domain logic for strict architectural validation but can also deploy that logic into a realistic, distributed network environment to measure the true operational overhead of these consistency models.

\subsection{The Gap in Current Methodologies and Tools}

Domain-Driven Design (DDD) and various transactional models are well-documented in isolation, but current solutions fail to provide an integrated environment where these elements can be experimentally validated before production. This gap is most evident in two primary areas:

\begin{enumerate}
\item While industry-standard frameworks like Eventuate Tram, Axon, and Temporal offer robust execution environments for distributed workflows, they operate primarily at the infrastructure level. As noted in the related work, these platforms lack native abstractions for DDD aggregates, forcing developers to manually map their domain models to the framework's coordination logic. Furthermore, because these tools require a full-scale deployment (databases, message brokers), they are unsuitable for the shift-left validation sought in RQ1. They do not allow architects to test the viability of domain boundaries or aggregate designs in a lightweight, deterministic manner before committing to a specific infrastructure.

\item Current distributed system simulators, such as SimBlock and VWR, or performance-centric tools like PerfSim and Terminus, are insufficient for answering RQ2 and RQ3. These tools treat microservices as "black boxes" focusing on network latency, resource consumption, or protocol-level exchanges while abstracting away the application-level logic. Therefore, they ignore the internal state transitions of DDD aggregates and the complex business rules that define system integrity. To effectively experiment with how domain-model boundaries and transactional guarantees (like Sagas vs. TCC) influence a system, a simulator must execute the actual domain logic rather than relying on statistical distributions.

\end{enumerate}

This work provides a simulator that directly incorporates DDD principles while remaining decoupled from communication and transactional infrastructure. Unlike existing tools, our approach enables a strict architectural validation of business logic, allowing developers to explore the trade-offs of their design decisions with minimal effort. Additionally, we intend to provide an environment where it is easy to switch between different transactional and communication models, which is addressed by RQ4.

\section{Microservice Simulator}
\label{sec:microservice-simulator}

The \textit{Microservice Simulator}\footnote{\url{https://github.com/socialsoftware/microservices-simulator}} provides a framework that enables developers to implement business logic using Domain-Driven Design (DDD) principles and subsequently execute that logic under configurable transactional models and network communication mechanisms.

\begin{figure*}[h]
    \centering
    \includegraphics[width=1\textwidth]{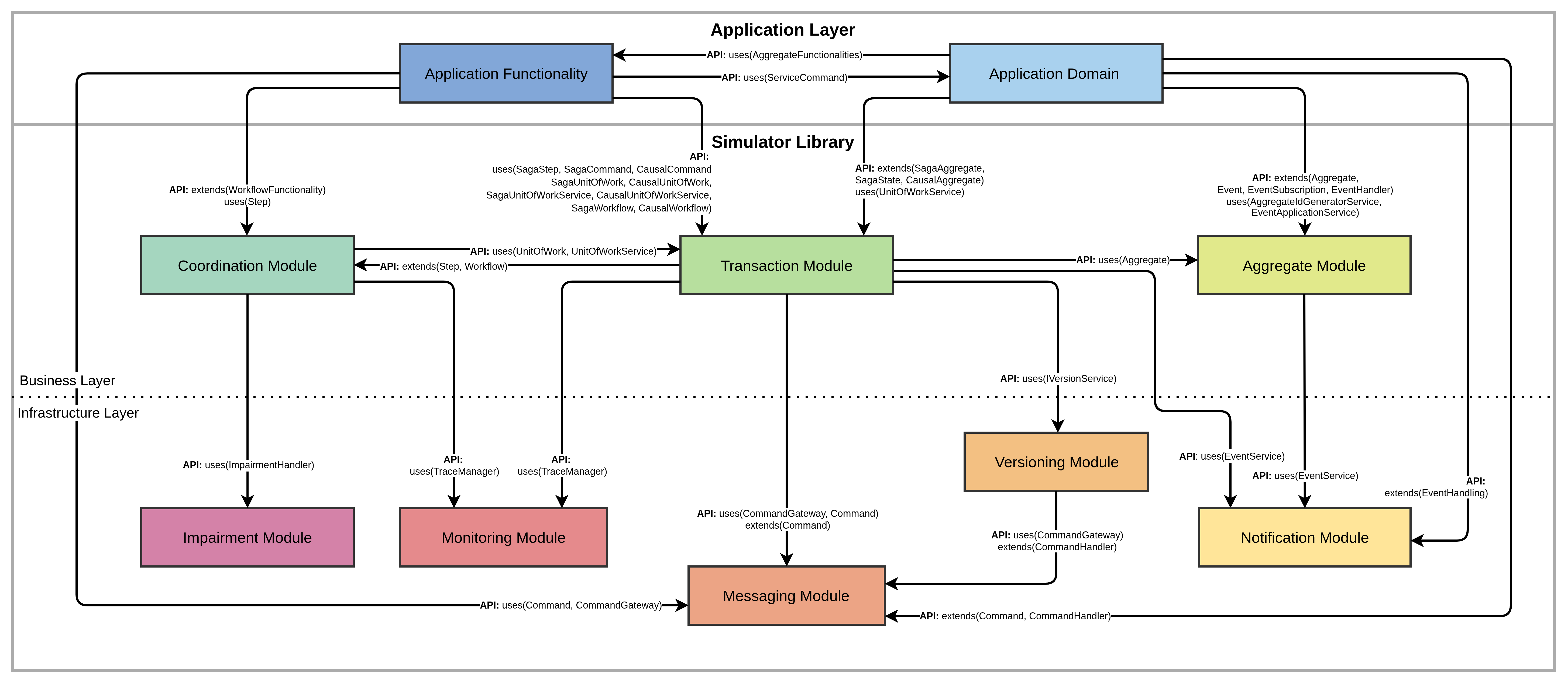}
    \caption{Structural architecture of the \textit{Microservice Simulator}. The \textit{Simulator Library} acts as a microservices domain-driven design middleware, explicitly decoupling the \textit{Application Layer's} domain logic from the underlying \textit{Business Layer} and \textit{Infrastructure Layer}.}
    \label{fig:simulator_layers}
\end{figure*}

As illustrated in Figure~\ref{fig:simulator_layers}, the system architecture is divided into three layers, ensuring a clear separation of concerns and well-defined module boundaries. The \textit{Application Layer} sits at the top, containing the concrete domain logic, specifically, the \textit{Application Functionality} and \textit{Application Domain} components. This layer is entirely decoupled from the underlying infrastructural complexities. 

Beneath the \textit{Application Layer} resides the \textit{Simulator Library}, which is further subdivided into two distinct tiers:
\begin{itemize}
    \item \textbf{Business Layer}: This middle tier provides the core coordination and domain structuring mechanisms. It encompasses the \textit{Coordination Module} (managing workflow execution), the \textit{Transaction Module} (supporting transaction models), and the \textit{Aggregate Module} (managing domain entity state and identity). The \textit{Application Layer} interacts with those layers through the exposed Application Programming Interfaces (APIs). This is done either by extending some base classes (e.g., inheriting from \texttt{Workflow Functionality} or \texttt{Aggregate}) or using some modules (e.g., use \texttt{Saga Step}).
    \item \textbf{Infrastructure Layer}: The foundational tier manages cross-cutting technical concerns and network operations. It includes the \textit{Messaging Module} (for direct inter-service communication), the \textit{Notification Module} (for event-based communication), the \textit{Impairment Module} (for performance delays and fault injection), the \textit{Monitoring Module} (for tracing), and the \textit{Versioning Module} (for unique ID generation).
\end{itemize}

The directional arrows in the architecture diagram denote strict API usage (\texttt{uses}) and inheritance (\texttt{extends}) relationships, enforcing a top-down dependency hierarchy.
The \textit{Application Layer} depends on the \textit{Simulator Library's Business Layer} to structure its logic, while the \textit{Business Layer} orchestrates the primitive operations provided by the \textit{Infrastructure Layer}. This modular architecture allows the simulator to swap underlying infrastructural implementations (e.g., changing synchronous gRPC messaging to asynchronous stream-based messaging) without requiring any modifications to the coordination and transactional models and the \textit{Application Layer's} domain logic.

\subsection{Messaging Module}
\label{subsec:messaging_module}

To achieve a realistic simulation of a distributed microservice architecture, the underlying communication infrastructure must be capable of adapting to various deployment topologies without intruding upon the application's domain logic. The \textit{Messaging Module} is specifically designed to address RQ3 by providing a decoupled communication backbone for downstream-to-upstream inter-aggregate interactions. By encapsulating local in-memory execution, remote procedure calls, and stream-based messaging behind a unified interface, this module enables developers to execute, benchmark, and evaluate the same business operations across fundamentally different network environments. Furthermore, this reliance on abstract interfaces establishes a highly extensible foundation, allowing architects to seamlessly integrate future or alternative communication protocols (e.g., REST, novel message brokers) without requiring any modifications to the core domain logic.

\begin{figure*}[t]
    \centering
    \includegraphics[width=0.9\textwidth]{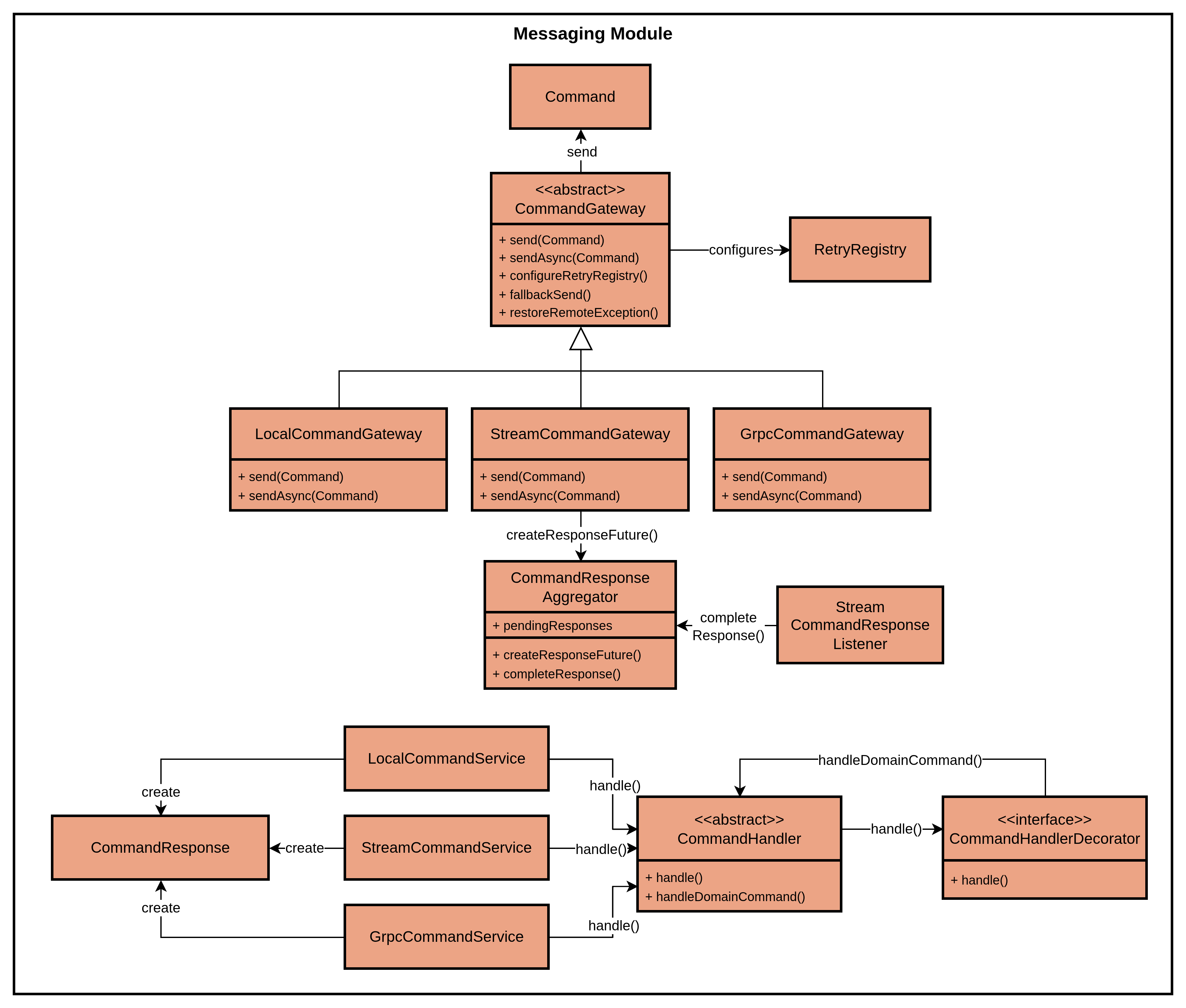}
    \caption{Detailed architecture of the \textit{Messaging Module}, illustrating the separation between command dispatching (\textit{Gateways}) and command receiving (\textit{Services} and \textit{Handlers}), along with their protocol-specific implementations.}
    \label{fig:simulator_layers_messaging}
\end{figure*}

The \textit{Messaging Module} supports direct communication between microservices in the \textit{Microservice Simulator}, enabling decoupled interactions between aggregates. From a strategic Domain-Driven Design perspective, this module is strictly responsible for \textbf{downstream-to-upstream communication}. It implements the \textit{Command} pattern~\cite{gamma1994design} to allow a downstream orchestrator to explicitly invoke upstream services, thereby abstracting the underlying transport protocols from the \textit{Application Layer}'s domain logic. The module is structured around two primary responsibilities: command dispatching and command handling.

\textbf{Command Dispatching:}
The dispatching mechanism is exposed to the Application Layer via the abstract \texttt{Command Gateway}. This gateway provides a standardized interface containing \texttt{send()} and \texttt{sendAsync()} methods, allowing the application to dispatch \texttt{Command} objects without knowing the destination's location or the transport protocol used. 

The asynchronous behavior is managed at the abstract level: the \texttt{sendAsync()} method simply wraps the synchronous \texttt{send()} execution within a \texttt{Completable Future} and offloads it to a managed thread pool. Consequently, when utilizing \texttt{sendAsync()}, it remains the strict responsibility of the application developer to handle the returned \texttt{Completable Future} (e.g., by chaining non-blocking callbacks or explicitly awaiting the result) if the subsequent business logic depends on the remote execution's outcome. 

To support various architectural topologies, the module provides three concrete implementations of the synchronous \texttt{send()} operation:
\begin{itemize}
    \item \textbf{\texttt{Local Command Gateway}}: Used in single-process deployments (such as testing environments). It dispatches commands directly in-memory, bypassing network serialization, executing the command synchronously on the caller's thread.
    \item \textbf{\texttt{Grpc Command Gateway}}: Facilitates high-performance remote procedure calls (RPC) using \textit{gRPC}. It is ideal for point-to-point communication where low latency is required. The \texttt{send()} method utilizes blocking gRPC stubs to halt the executing thread until a response is received from the remote server.
    \item \textbf{\texttt{Stream Command Gateway}}: Utilizes event-streaming platforms (e.g., Kafka or RabbitMQ) for decoupled, message-driven command dispatching. Because streaming is inherently asynchronous, the gateway relies on a correlation ID mechanism. To provide the synchronous \texttt{send()} API over this medium, the gateway publishes the command, generates a \texttt{Completable Future} tied to the correlation ID via a \texttt{Command Response Aggregator}, and explicitly blocks the thread until the corresponding response is received on a dedicated reply channel or a timeout occurs.
\end{itemize}

\textbf{Command Handling:}
On the receiving end, incoming commands are processed by protocol-specific command services (\texttt{Local Command Service}, \texttt{Grpc Command Service}, and \texttt{Stream Command Service}). These infrastructure-level services are responsible for deserializing the incoming requests, invoking the appropriate application logic, and wrapping the result into a \texttt{Command Response}.

The execution of the business logic itself is delegated to implementations of the abstract \texttt{Command Handler} class. The \texttt{Command Handler} utilizes a \textit{Template Method} pattern~\cite{gamma1994design}: its public \texttt{handle()} method manages cross-cutting concerns (such as serializable transaction isolation) through the application of the Decorator pattern~\cite{gamma1994design} (\texttt{Command Handler Decorator}), while delegating the actual domain logic execution to the abstract \texttt{handleDomainCommand()} method, which concrete domain handlers must implement.

The \texttt{Command Handler Decorator} interface provides a crucial extensibility point within the messaging pipeline. By applying the \textit{Decorator} pattern, it allows higher-level middleware (such as distributed transaction coordinators) to be transparently injected into the command execution flow. This design ensures separation of concerns, allowing the \textit{Messaging Module} to route commands without needing any knowledge of the overarching business or transactional context.

\textbf{Resilience and Fault Tolerance (Circuit Breaker \& Retries):}
Because distributed systems are inherently prone to transient failures, such as network timeouts, temporary service unavailability, or optimistic concurrency conflicts, the \textit{Messaging Module} incorporates a resilient retry mechanism. Utilizing the Resilience4j framework, the \texttt{Command Gateway} wraps its synchronous send operation with an active \texttt{@Retry} circuit breaker (e.g., explicitly configured via a \texttt{Retry Registry}). When a command fails due to a recoverable infrastructure or concurrency exception (often encapsulated as a generic \texttt{Runtime Exception} over the wire), the gateway automatically intercepts the failure and applies an exponential backoff retry strategy. This also extends to asynchronous communication, as the \texttt{sendAsync} method simply executes the \texttt{send} operation in a background thread, guaranteeing that retries are always applied.

To ensure domain logic remains unaffected by transient networking issues, the abstract \texttt{Command Gateway} employs a custom exception deserialization strategy. During remote invocations, specific errors are flattened into generic network exceptions. To prevent the circuit breaker from endlessly retrying fatal business rule violations, the gateway uses Java Reflection to reconstruct the original exception object from its class name and message payload. If the reconstructed error is identified as a domain-level \texttt{Simulator Exception}, it bypasses the retry loop and immediately aborts the workflow. Conversely, standard infrastructure or concurrency errors trigger the exponential backoff retry strategy. Finally, a dedicated \texttt{fallbackSend()} method acts as the ultimate safety net; if all retry attempts are exhausted, it throws a structured exception to cleanly notify the orchestrator of the remote service's unavailability.

\subsection{Notification Module}
\label{subsec:notification_module}

To directly address RQ1 (facilitating DDD experimentation), the simulator must provide a transparent mechanism for the implementation of upstream-to-downstream relationship communication through asynchronous event propagation. The \textit{Notification Module} serves this exact purpose by abstracting the complexities of event persistence, polling, and message broker integration. This module ensures that developers can utilize domain events without coupling their business logic to the underlying infrastructure (RQ3). Consequently, the simulator can seamlessly transition the event-processing model from a local, centralized database to a fully distributed, stream-based network without requiring any modifications to the application code.

\begin{figure*}[t]
 \centering
 \includegraphics[width=0.6\textwidth]{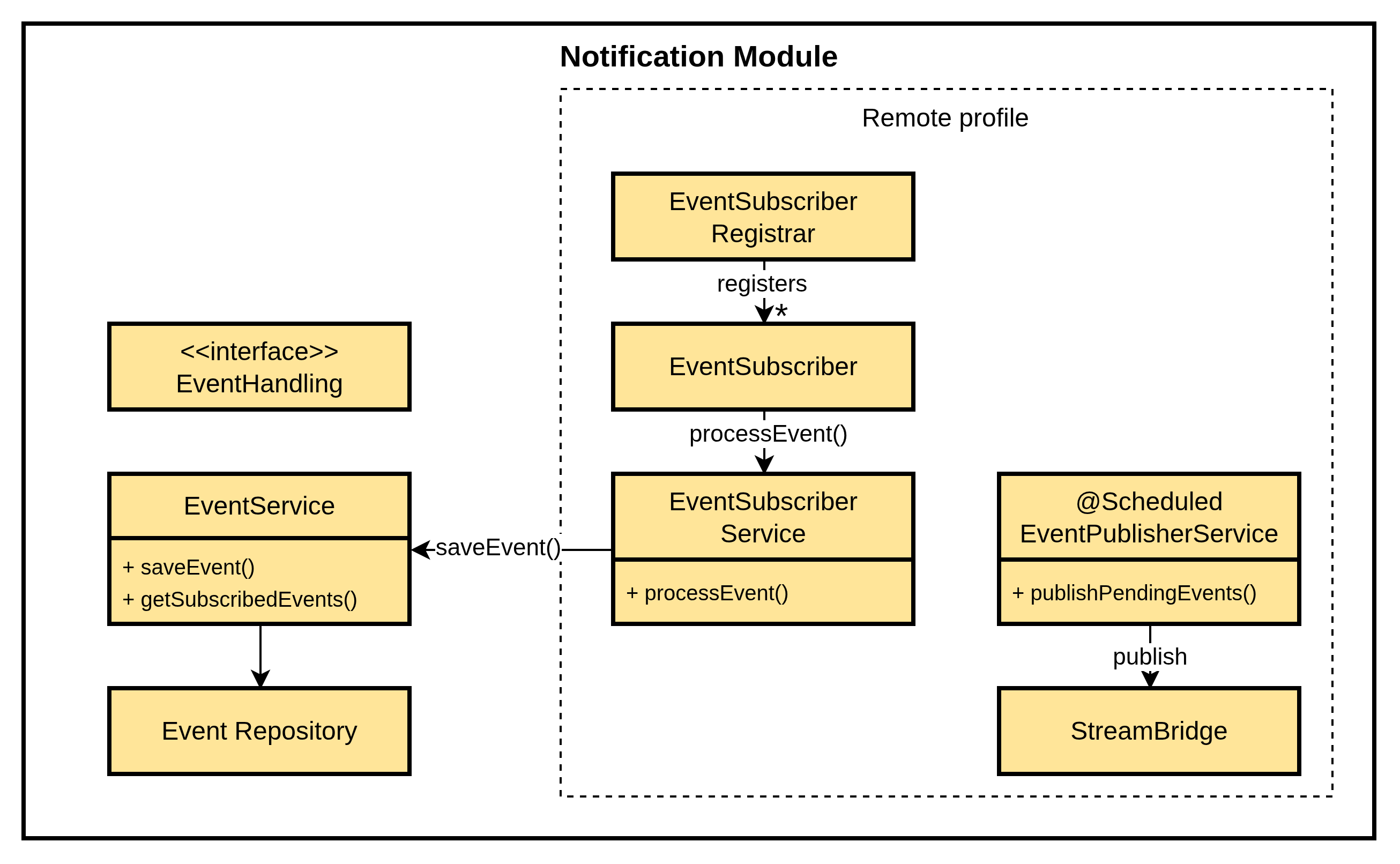}
 \caption{Architecture of the \textit{Notification Module}, highlighting the remote-profile event publication/subscription path and the local event persistence services.}
 \label{fig:simulator_layers_notification}
\end{figure*}

The \textit{Notification Module} provides the infrastructure required for event communication between microservices (aggregates). As shown in Figure~\ref{fig:simulator_layers_notification}, its core persistence path is centered on the \texttt{Event Service} and \texttt{Event Repository}, which reliably maintain the event log and expose the queries necessary for downstream event processing. The \texttt{Event Service} interface supports the publishing (\texttt{saveEvent()} and subscription (\texttt{getSubscribed()}) of events.

The implementation differs depending on whether the deployment is local or distributed. In a local deployment, there is a single \texttt{Event Repository} and all aggregates publish and subscribe to the same database. In the distributed deployment, there is a \texttt{Event Service} and \texttt{Event Repository} for each microservice (aggregate), and it is necessary to use a communication infrastructure to propagate the events between microservices (repositories). Each microservice publishes its events in its repository, and they are propagated to the repositories of the subscriber microservices.

In Figure~\ref{fig:simulator_layers_notification}, the remote profile highlights the \textit{Notification Module} elements that are necessary for the distributed deployment. The \texttt{Event Publisher Service} operates on a scheduled execution loop (\texttt{@Scheduled}), periodically polling the local database for unpublished events, serializing the payloads, and dispatching them via a \texttt{Stream Bridge} to the unified \texttt{event-channel}. Upon successful delivery to the broker, the local records are atomically marked as published. On the receiving end, the \texttt{Event Subscriber Registrar} reads the application configuration to dynamically create and register stream consumer beans (\texttt{EventSubscriber}s). These dynamic beans act as the direct listeners to the external message broker. Upon receiving an incoming message, the generated subscriber delegates the payload to the \texttt{Event Subscriber Service}. Only explicitly subscribed event types are then deserialized and durably stored in the microservice's local \texttt{Event Repository} via the \texttt{Event Service}.

Ultimately, the Notification Module strictly encapsulates transport, persistence, and scheduling boundaries through the \texttt{Event Handling} contract. Within each microservice, concrete scheduled handlers trigger local event consumption cycles, but the actual domain-level subscription resolution and per-event dispatch are delegated entirely to the \textit{Aggregate Module} (\texttt{Event Application Service} and \texttt{Event Handler}). This strict architectural segregation ensures that event publication, network transport, and durable storage remain fully independent from the application's domain, preserving a uniform event-processing model across both local and distributed deployments.

\subsection{Impairment Module}
\label{subsec:impairment_module}

A critical advantage of the shift-left methodology in microservice architecture is the ability to validate complex coordination logic before a full-scale production deployment. To fully address RQ2 (experimenting with diverse transactional models), a simulator must empirically evaluate how protocols like Sagas and TCC recover from adverse conditions. Furthermore, to address RQ4 (minimizing developer effort), the simulator must automate the generation of these complex failure interleavings. The \textit{Impairment Module} is an infrastructure component dedicated to fulfilling these objectives by testing the performance delays and fault-tolerance mechanisms of the \textit{Microservice Simulator}. It provides the capability to deterministically inject faults, such as microservice crashes, and performance delays, into the execution flows of microservice systems. By systematically injecting these impairments, developers and researchers can rigorously evaluate the application's behavior under such conditions.

\begin{figure*}[t]
    \centering
    \includegraphics[width=0.45\textwidth]{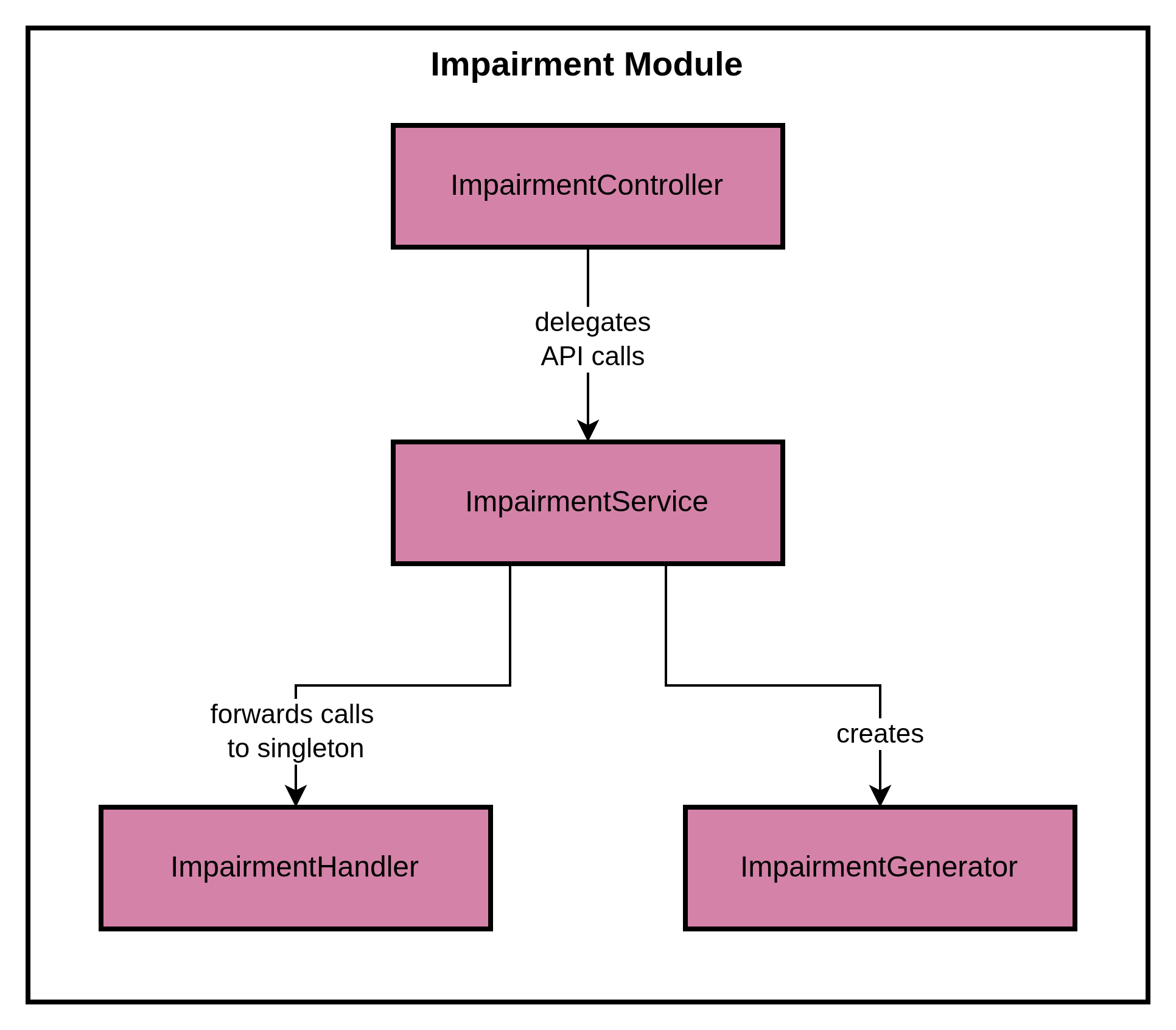}
    \caption{Architecture of the \textit{Impairment Module}, illustrating the controller-service delegation and the interaction with the \texttt{Impairment Generator} and the \texttt{Impairment Handler} singleton for testing performance delays and fault tolerance.}
    \label{fig:simulator_layers_impairment}
\end{figure*}

As illustrated in Figure~\ref{fig:simulator_layers_impairment}, the module is composed of four main elements, structured to separate API interactions from the core fault-injection logic:

\begin{itemize}
    \item \textbf{\texttt{Impairment Controller}}: This component exposes RESTful API endpoints (such as \texttt{/load} and \texttt{/clean}) to orchestrate the testing environment. It acts as the HTTP entry point for external test scripts to trigger the loading of new fault behavior profiles or to clean up state between test runs. It delegates all operational requests directly to the \texttt{Impairment Service}.
    
    \item \textbf{\texttt{Impairment Service}}: Acting as a coordinating intermediary, this service abstracts the underlying logic from the controller. When instructed to load a new impairment scenario, the service determines the correct directory paths within the testing environment and configures the \texttt{Impairment Handler}. It is also responsible for invoking the \texttt{Impairment Generator} when automated test behavior combinations need to be created.
    
    \item \textbf{\texttt{Impairment Generator}}: This component is responsible for parsing high-level test definition files and generating exhaustive, balanced combinations of failure behaviors across multiple functionalities. It calculates cross-products of specified faults and delays, ultimately producing detailed CSV configuration files (e.g., one per functionality) that explicitly define when and where faults should occur during a test run.
    
    \item \textbf{\texttt{Impairment Handler}}: The core execution engine of the module, implemented as a thread-safe Singleton. During application execution, domain steps or transactional coordination mechanisms consult the \texttt{Impairment Handler} to determine if an impairment should be triggered. The handler parses the CSV files generated by the \texttt{Impairment Generator} on demand, maintains internal counters to track execution iterations (e.g., to simulate a fault only on the first attempt but succeed on a retry), and manages test reporting by logging impairment events to a dedicated report file.
\end{itemize}

In centralized deployments, the \texttt{Impairment Handler} operates as a simple, thread-safe Singleton. However, in distributed topologies, each microservice runs in an isolated Java Virtual Machine (JVM). Therefore, a fault configuration loaded into one microservice is not visible to the others, making it difficult to simulate complex, cross-service failure scenarios.

To solve this distributed configuration problem, the simulator utilizes a broadcast mechanism within the \textit{API Gateway}. When a testing tool (such as a JMeter script) initiates a test, it sends a single HTTP \texttt{POST} request to the gateway, providing the directory of the impairment configuration files. Upon receiving this request, the gateway broadcasts the load request to every microservice via a REST client.

Each independent microservice receives this broadcast at its local \texttt{Impairment Controller}, which then configures its own local \texttt{Impairment Handler} singleton to point to the shared test configuration files. In the simulator, fault injection is managed by the \texttt{Execution Plan}, which runs within the orchestrating microservice of a functionality. Therefore, as a distributed transaction (such as a Saga) executes, the orchestrator queries its local handler to inject delays or exceptions into the workflow steps.

\subsection{Monitoring Module}
\label{subsec:monitoring_module}

A fundamental requirement for evaluating and debugging any distributed architecture is comprehensive observability. To effectively address RQ2 and RQ3, researchers and developers must be able to visualize the performance, network latency, and execution bottlenecks of complex coordination, such as Sagas and TCC. Furthermore, to satisfy RQ4 (minimizing developer effort), this observability infrastructure must be seamlessly integrated into the simulator, automatically propagating execution context without polluting the pure domain logic. The \textit{Monitoring Module} provides the necessary infrastructure to observe and trace functionalities and commands as they flow across various microservices in the \textit{Microservice Simulator}. It implements tracing to seamlessly correlate operations, even when they cross network boundaries or process asynchronous messages. The module relies on the OpenTelemetry (OTel) framework to collect and export trace data to external observability backends (e.g., Jaeger).

\begin{figure*}[t]
    \centering
    \includegraphics[width=0.25\textwidth]{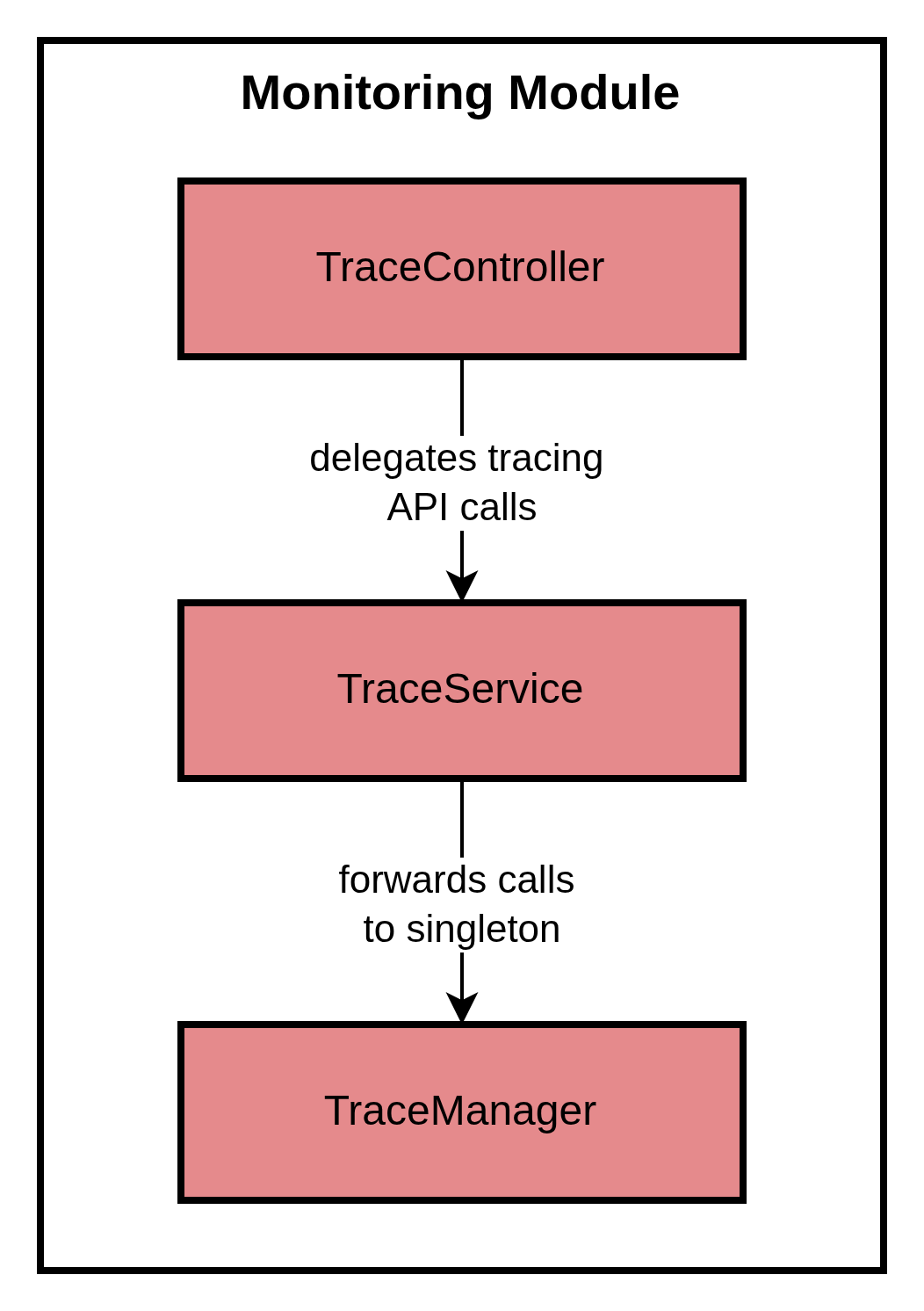}
    \caption{Architecture of the \textit{Monitoring Module}, depicting the controller-service-singleton pattern used to manage tracing across the system.}
    \label{fig:simulator_layers_monitoring}
\end{figure*}

As depicted in Figure~\ref{fig:simulator_layers_monitoring}, the module is divided into three key components:

\begin{itemize}
    \item \textbf{\texttt{Trace Controller}}: This component exposes a set of RESTful API endpoints (e.g., \texttt{/createRoot}, \texttt{/start}, \texttt{/end}, \texttt{/flush}) that allow external testing frameworks or orchestration scripts to explicitly initiate, terminate, and flush traces. Rather than containing the tracing logic itself, the controller acts purely as an HTTP entry point, delegating all incoming requests to the \texttt{Trace Service}.

    \item \textbf{\texttt{Trace Service}}: This service acts as an intermediary orchestration layer. It receives the tracing API calls from the \texttt{Trace Controller} and forwards them to the underlying \texttt{Trace Manager}. Furthermore, it binds the module to the Spring application's environment (e.g., by automatically configuring the local \texttt{service.name} during the application's initialization phase) before interacting with the singleton manager.

    \item \textbf{\texttt{Trace Manager}}: This is a thread-safe Singleton class responsible for the core lifecycle management of OpenTelemetry spans. It configures the \texttt{Open Telemetry SDK} and initializes the \texttt{OTLP gRPC Span Exporter}, which is responsible for exporting the gathered trace data to an external collector via \textit{gRPC}. The manager handles the creation of \textit{master root spans} to encapsulate entire coordination workflows, as well as local \textit{root spans} for individual microservices. Crucially, it provides mechanisms to inject and extract remote parent contexts (Trace IDs and Span IDs), allowing disparate microservices to correctly link their local traces to the broader coordination tree.
\end{itemize}

In distributed deployments, the simulator ensures trace synchronization via a coordination mechanism in the \textit{API Gateway}. When an external test script (e.g., JMeter) triggers the \texttt{/traces/start} endpoint on the gateway, the gateway's \texttt{Admin Controller} performs a two-stage synchronization. First, it requests a master context (Trace ID and Span ID) from one of the active microservices via the \texttt{/traces/createRoot} endpoint. Second, it broadcasts this master context to all discovered microservices in the cluster via their local \texttt{Trace Controller}. This synchronized start ensures that every independent node links its local execution spans to the same global coordination tree, enabling a unified, end-to-end visualization of distributed transactions in the observability backend.

\subsection{Versioning Module}
\label{subsec:versioning_module}

To effectively address RQ2 (experimenting with diverse transactional models), the implementation of transactional models within the simulator relies heavily on optimistic concurrency control to manage distributed state changes. Consequently, it is strictly necessary to assign and track explicit versions of the domain aggregates to accurately detect and resolve concurrent modifications. However, establishing a robust versioning system is not a trivial task and introduces significant architectural trade-offs. To satisfy RQ4 (minimizing developer effort), the \textit{Versioning Module} completely abstracts the immense complexity of these trade-offs from the application developer. On one hand, relying on a centralized service for the generation of unique identifiers can easily become a severe performance bottleneck under high load due to network overhead and database lock contention. On the other hand, the generation of distributed unique identifiers without a central coordinator is a fundamentally hard problem in distributed systems, requiring complex algorithms to guarantee global uniqueness, prevent cross-node collisions, and preserve logical ordering.

\begin{figure*}[t]
    \centering
    \includegraphics[width=0.9\textwidth]{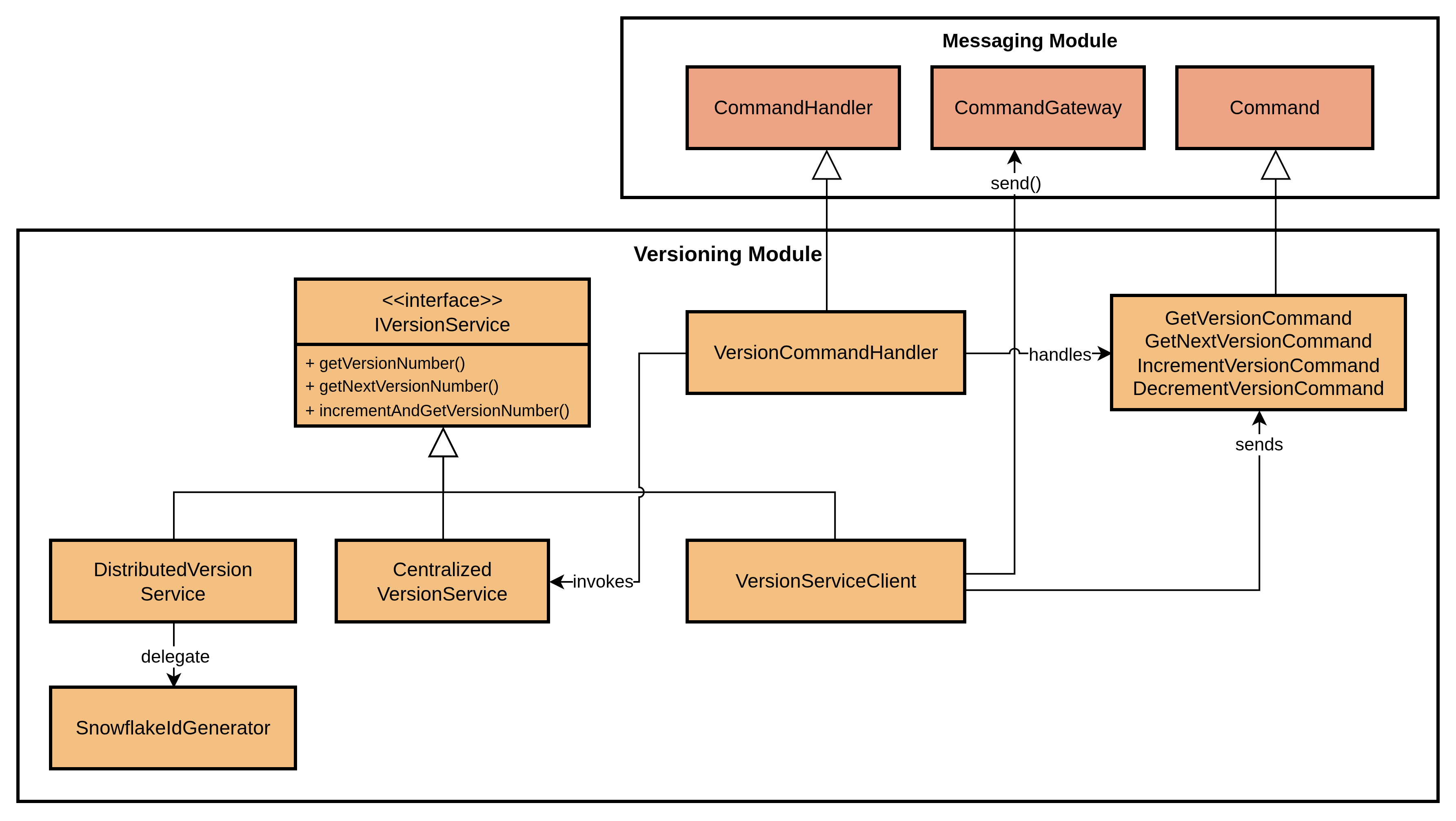}
    \caption{Detailed architecture of the \textit{Versioning Module}, illustrating the \texttt{IVersion Service} interface and its three specific implementations, alongside the remote command dispatching mechanism integrated with the \textit{Messaging Module}.}
    \label{fig:simulator_layers_versioning}
\end{figure*}

The \textit{Versioning Module} is an infrastructure component responsible for generating and managing version numbers across the system. It exposes its functionality through the \texttt{IVersion Service} interface, which is primarily consumed by the \textit{Transaction Module} to assign unique versions to domain entities and transactions. The \texttt{IVersion Service} defines core operations for version retrieval and manipulation, namely \texttt{getVersionNumber()}, \texttt{getNextVersionNumber()}, and \texttt{incrementAndGetVersionNumber()}.

To accommodate different architectural topologies and scalability requirements, the module provides three distinct implementations of the \texttt{IVersion Service}:

\begin{itemize}
    \item \textbf{Centralized Version Service}: This implementation manages version numbers in a centralized manner. It relies on a persistent storage mechanism to maintain the current version state. When a new version is requested, the service retrieves the current value, performs a sequential increment ($+1$), and updates the repository. While straightforward, this approach can become a bottleneck in highly concurrent environments due to database lock contention.
    
    \item \textbf{Distributed Version Service}: Designed for highly concurrent and distributed deployments, this service generates unique version identifiers without the overhead of a centralized repository. It achieves this by delegating ID creation to a \texttt{Snowflake Id Generator}~\cite{snowflake}. Based on the algorithm originally developed by \textit{Twitter}, it generates unique, roughly time-ordered 64-bit integers by combining a custom epoch timestamp, a unique machine identifier, and a per-millisecond sequence number. This decentralized approach avoids cross-node collisions and ensures high availability, horizontal scalability, and minimal latency.
    
    \item \textbf{Version Service Client}: This implementation acts as a remote proxy, enabling application instances to interact with a standalone \texttt{Centralized Versioning Service} over the network. It integrates deeply with the Messaging Module by utilizing the \texttt{Command Gateway} to dispatch versioning requests as commands. Specifically, it sends \texttt{Get Version Command}, \texttt{Get Next Version Command}, \texttt{Increment Version Command}, and \texttt{Decrement Version Command}. On the receiving end, a \texttt{Version Command Handler} intercepts these commands and invokes the corresponding local operations on the underlying \texttt{Centralized Version Service} instance, which processes the request and synchronously routes the generated version ID back to the calling microservice.
\end{itemize}

\subsection{Aggregate Module}
\label{subsec:aggregate_module}

To directly address RQ1 (facilitating DDD experimentation), the simulator must provide native constructs that allow developers to accurately model the business logic and structural boundaries of their domains. Furthermore, to address RQ4 (minimizing developer effort), these domain models must remain isolated and completely decoupled from infrastructural complexities. Additionally, they should encapsulate all abstract DDD logic, ensuring that any extensions only need to define the concrete implementation. The \textit{Aggregate Module} serves as this layer, providing the core abstractions necessary to enforce internal data consistency, manage entity identity, and explicitly declare upstream-downstream dependencies.

\begin{figure*}[t]
 \centering
 \includegraphics[width=0.45\textwidth]{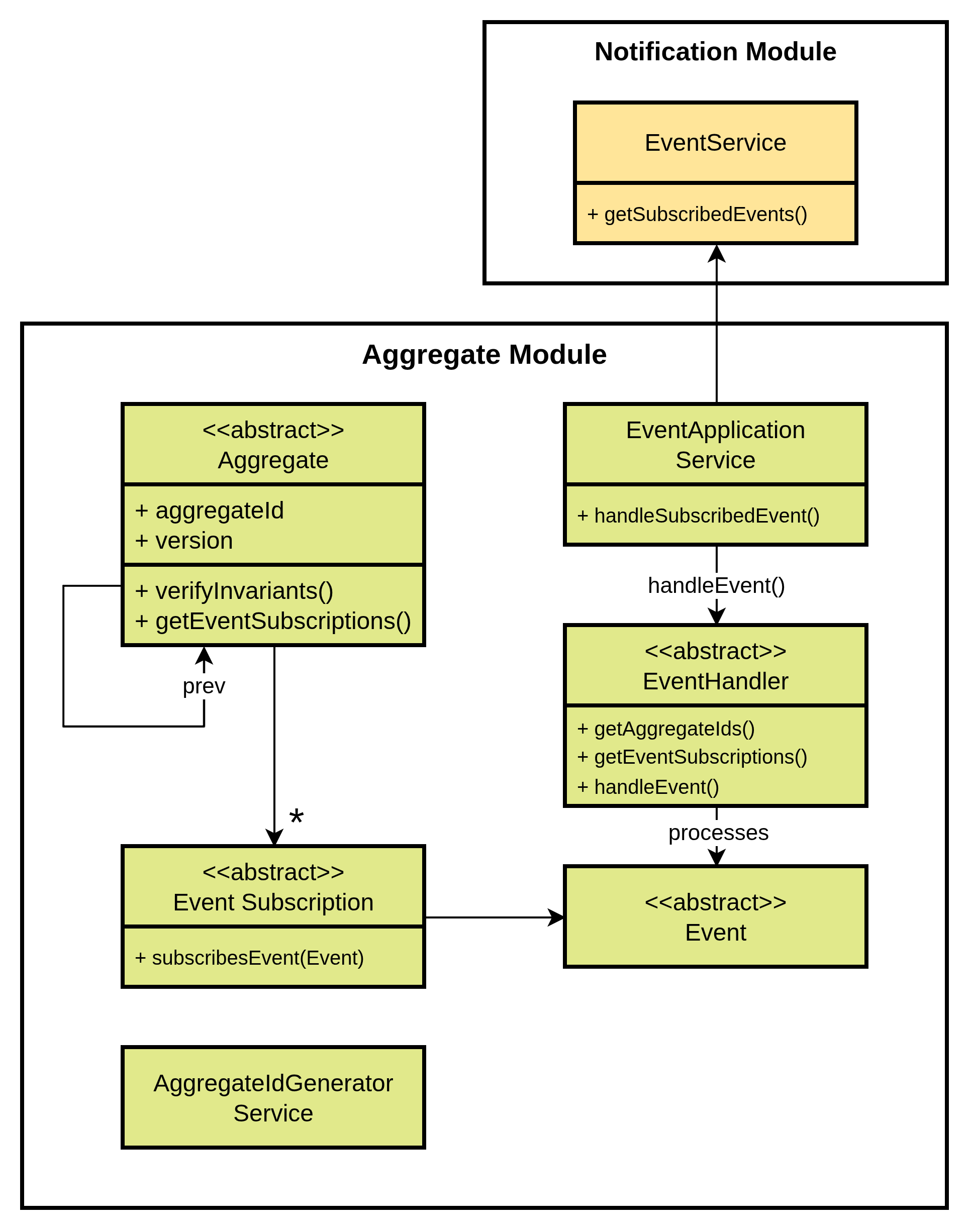}
 \caption{Architecture of the \textit{Aggregate Module}, showing the core \texttt{Aggregate} abstraction, identity generation services, event subscription model, and integration points with event handling.}
 \label{fig:simulator_layers_aggregate}
\end{figure*}

The \textit{Aggregate Module} serves as the simulator's core unit of consistency and versioned state evolution. Its central abstraction is the \texttt{Aggregate} class, which encapsulates a logical identifier (\texttt{aggregateId}), an aggregate version sequence (\texttt{version}), lifecycle state, type metadata, and a \texttt{prev} pointer referencing its predecessor. This structure facilitates a copy-on-write state evolution model, where each successfully committed mutation generates a new, immutable aggregate version while preserving the historical state chain. Furthermore, the abstraction strictly enforces domain responsibilities by mandating the implementation of \texttt{verifyInvariants()} and \texttt{getEventSubscriptions()} methods (for intra-aggregate consistency validation and declaring upstream-to-downstream inter-aggregate dependencies, respectively).

Aggregate identity management is completely decoupled from versioning and is governed by the \texttt{Aggregate Id Generator Service}. This service allocates unique logical \texttt{aggregateId} values upon instantiation. By segregating logical identity generation from version assignment, which is deferred to the Transaction Module at commit time, the simulator ensures that identity remains deterministic and consistent regardless of the underlying distributed concurrency protocol.

Finally, the module encapsulates the event-domain abstractions that realize the upstream-to-downstream relationships among aggregates: the abstract \texttt{Event}, \texttt{Event Handler}, \texttt{Event Subscription}, and \texttt{Event Application Service}. It is crucial to distinguish these domain-level definitions from the infrastructural \textit{Notification Module}. Aggregates explicitly declare their subscriptions to changes in upstream aggregates via \texttt{getEventSubscriptions()}, while the \texttt{Event Handler} dynamically resolves subscriber IDs and matches subscriptions to incoming event types. During execution, the \texttt{Event Application Service} queries the underlying \textit{Notification Module} (via \texttt{getSubscribedEvents()}) and delegates the retrieved payloads to the specific domain handler logic. Consequently, the \textit{Aggregate Module} acts as both the strict state boundary for direct communication (associated with downstream-to-upstream communication) and the contract boundary for asynchronous event propagation (associated with upstream-to-downstream communication). The former is only explicit in the \textit{Application Layer}, where the \textit{Command} and \textit{CommandHandler} of infrastructural \textit{Messaging Module} are extended to define the aggregate interfaces. Concerning the latter, it is delegated to the \textit{Notification Module} the persistence and routing of aggregate's events.

\subsection{Coordination Module}
\label{subsec:coordination_module}

To address RQ1 (facilitating DDD experimentation) and RQ2 (experimenting with diverse transactional models), the simulator requires a robust orchestration engine capable of managing distributed workflows. Furthermore, to satisfy RQ4 (minimizing developer effort), this orchestration must completely abstract boilerplate complexities away from the domain developer.

The \textit{Coordination Module} is a core module that dictates how complex, multi-step application functionalities are orchestrated and executed. It acts as the bridge between the \textit{Application Layer}'s domain logic and the underlying infrastructural services (\textit{Transaction}, \textit{Monitoring}, and \textit{Impairment}). Additionally, by abstracting the execution flow into explicitly defined workflows and execution plans, the module allows the simulator to inject observability and deterministic failure behaviors uniformly across all functionalities.

\begin{figure*}[t]
    \centering
    \includegraphics[width=0.6\textwidth]{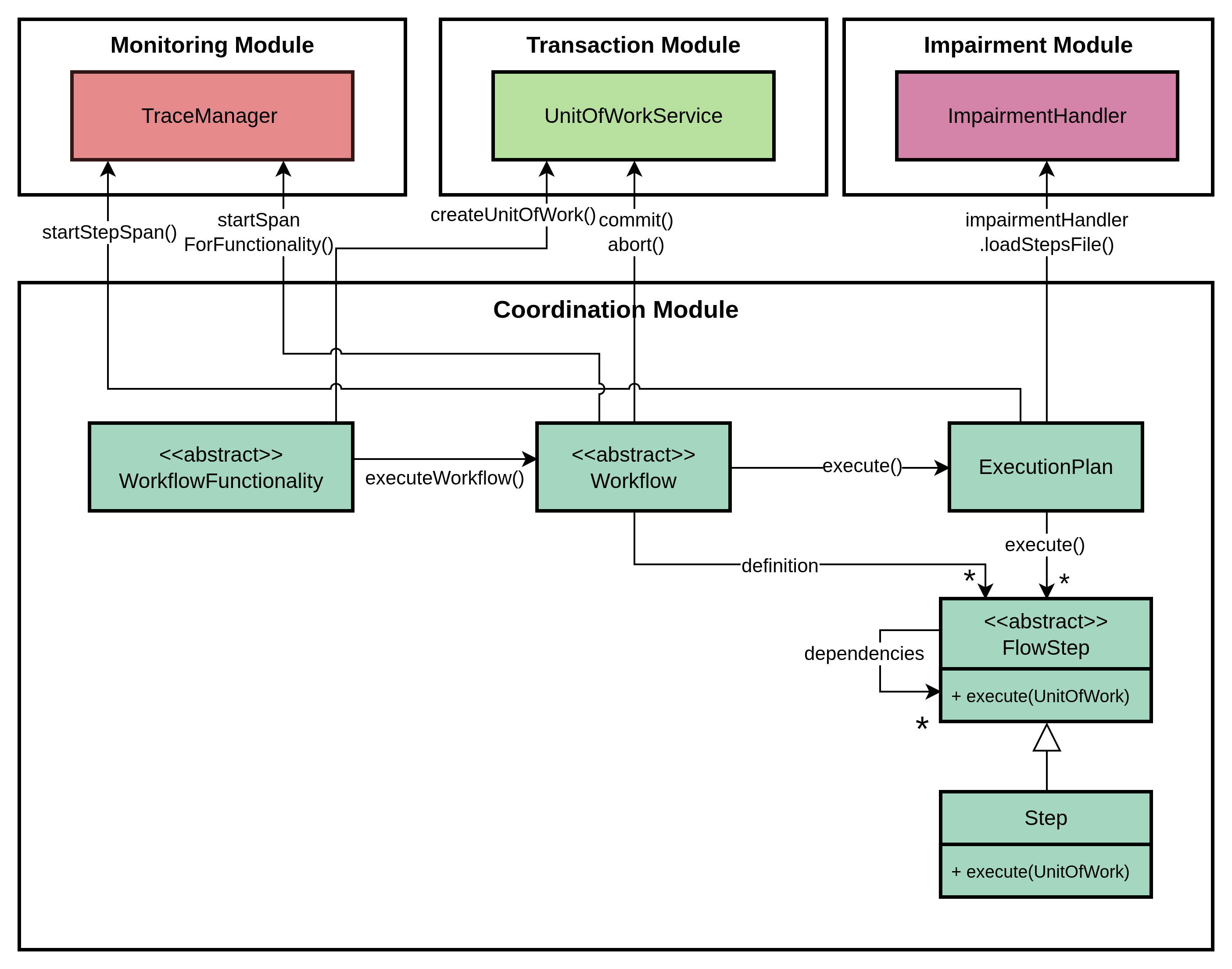}
    \caption{Architecture of the \textit{Coordination Module}, highlighting the components responsible for structuring, scheduling, and executing domain functionality workflows and their integration with the \textit{Transaction}, \textit{Monitoring}, and \textit{Impairment} modules.}
    \label{fig:simulator_layers_coordination}
\end{figure*}

The module is structured around several key classes that govern the execution lifecycle:

\begin{itemize}
    \item \textbf{\texttt{Workflow Functionality}}: This is the primary abstraction for any high-level domain operation that requires the coordination of several microservices. It encapsulates a specific \texttt{Workflow} and provides the public \texttt{executeWorkflow()} method, which initializes the execution process. By inheriting from this class within the \textit{Application Layer} and performing the required instantiations, functionalities automatically integrate with the simulator's transaction and execution mechanisms.
    
    \item \textbf{\texttt{Workflow}}: The \texttt{Workflow} class serves as the coordinating engine for a functionality. It holds references to all the constituent \texttt{Flow Step} objects and their respective execution dependencies. It is responsible for initiating the \texttt{Execution Plan} and handling the overarching transaction lifecycle via the \texttt{Unit Of Work Service} of the \textit{Transaction Module} (invoking \texttt{commit()} upon success or \texttt{abort()} upon failure). The \texttt{Workflow} class also interacts with the \textit{Monitoring Module} (\texttt{Trace Manager}) to explicitly demarcate the boundaries (start and end spans) of the functionality's execution.
    
    \item \textbf{\texttt{Execution Plan}}: The \texttt{Execution Plan} is responsible for the actual scheduling the workflow steps. It takes the raw steps and their dependency graphs from the \texttt{Workflow} and determines the correct execution order. During execution, it heavily integrates with the \textit{Impairment Module} (\texttt{Impairment Handler}). Before executing a step, the \texttt{Execution Plan} queries the \texttt{Impairment Handler} to determine if a specific behavior (e.g., a forced delay or a thrown \texttt{Simulator Exception}) has been scheduled for that precise step. It also coordinates with the \texttt{Trace Manager} to ensure these steps and delays are properly traced.
    
    \item \textbf{\texttt{Flow Step}} and \textbf{\texttt{Step}}: The \texttt{Flow Step} is an abstract representation of a discrete execution step within a larger functionality. It defines an \texttt{execute(UnitOfWork)} method and holds a list of its dependency steps. The concrete \texttt{Step} class provides the actual implementation, typically wrapping a domain operation (passed as a Java \texttt{Runnable}). Because steps explicitly declare their dependencies, the \texttt{Execution Plan} can guarantee that a step is only invoked once all its prerequisite steps have successfully completed.
\end{itemize}

\subsection{Transaction Module}
\label{subsec:transaction_module}

To effectively address RQ2 (evaluating diverse transactional models), the simulator requires a specialized infrastructure capable of enforcing distinct distributed transactional models, specifically orchestrating Sagas and Transactional Causal Consistency (TCC). To satisfy RQ4 (minimizing developer effort), the complexities of these coordination protocols must be completely abstracted from the application's business rules. The \textit{Transaction Module} is the central component dedicated to fulfilling these requirements, maintaining data consistency across distributed boundaries by providing interchangeable, simulator-level transaction management without polluting the core domain logic.

\begin{figure*}[htbp]
    \centering
    \includegraphics[angle=90, height=0.85\textheight]{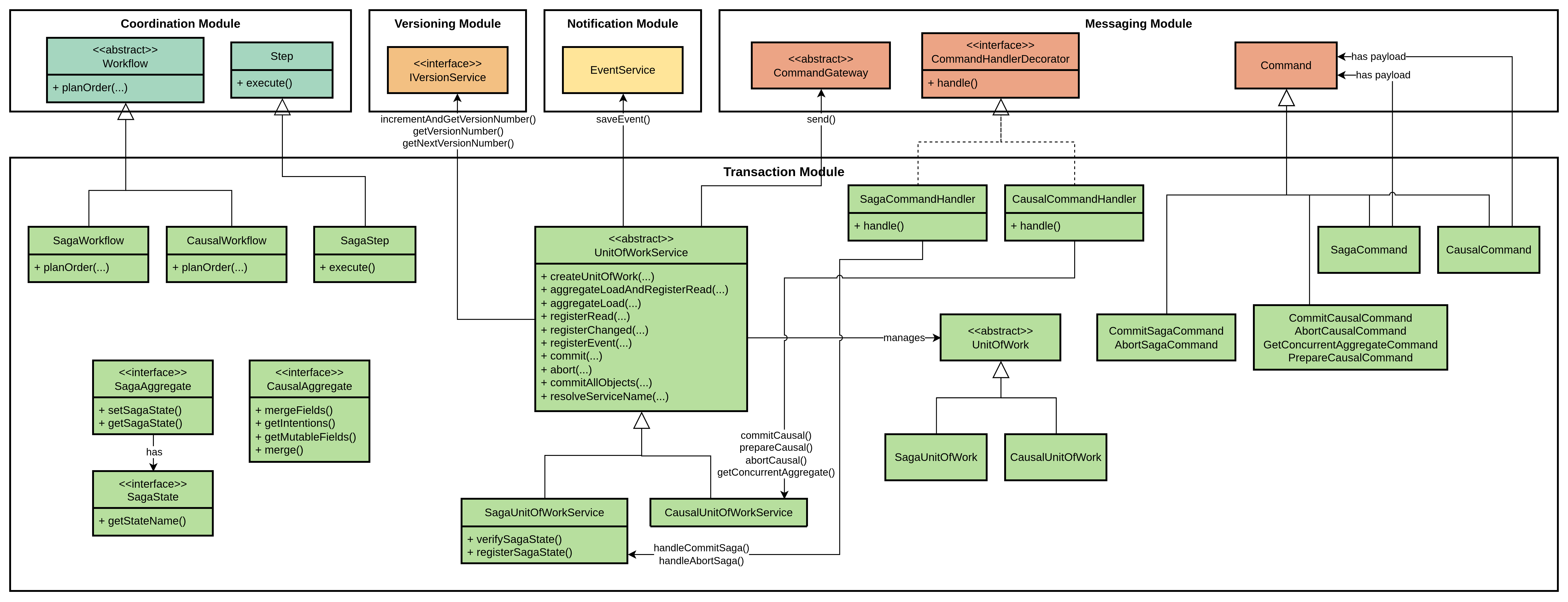}
    
    \caption{Architecture of the \textit{Transaction Module}, detailing the required extensions to the \textit{Aggregate}, the application of the \textit{Unit of Work} pattern, and the extension of the \textit{Messaging Module} to add transactional qualities to the microservices invocations.}
    \label{fig:simulator_layers_transaction}
\end{figure*}

The \textit{Transaction Module} is a central component responsible for maintaining data consistency across distributed boundaries. To support the exploration of various coordination protocols, this module explicitly abstracts transaction management away from the core domain logic to provide interchangeable mechanisms, specifically orchestrating Sagas and Transactional Causal Consistency (TCC).

The module's architecture is anchored by the \texttt{Unit Of Work} pattern~\cite{fowler03}, which formally isolates transaction boundaries and tracks state mutations. The primary components of this module include:

\begin{itemize}
 \item \textbf{\texttt{Unit Of Work Service}}: This abstract service acts as the primary entry point for the Application Layer's \texttt{Workflow}s. It exposes primitive operations for managing transactions, which include the definition of the transaction boundaries, through \texttt{createUnitOfWork}, \texttt{commit} and \texttt{abort}, and the registration of all accesses and changes to aggregates and events that are emitted, through methods such as \texttt{aggregateLoad}, \texttt{registerChanged}, and \texttt{registerEvent}. It utilizes the \textit{Versioning Module} to associate version numbers with aggregates, supporting invariants that refer to the previous state, and to support optimistic concurrency control for the \textit{Transactional Causal Consistency} (TCC) model. 
 Additionally, it uses \texttt{Event Service} of the \textit{Notification Module} to  publish the events emitted during the functionality execution. By atomically writing the events together with the aggregate changes, it applies the \textit{Transactional Outbox} pattern~\cite{richardson19} to guarantee atomicity of aggregate change and event emission.
 
 \item \textbf{Protocol-Specific Services (\texttt{Saga Unit Of Work Service} \& \texttt{Causal Unit Of Work Service})}: The two transactional models currently supported by the \textit{Microservice Simulator} are managed through two implementations of the \texttt{Unit Of Work Service}:
 \begin{itemize}
     \item \textbf{\texttt{Saga Unit Of Work Service}}: It manages semantic locks for atomic transactions across microservice aggregates to compensate for the lack of isolation. The implementation of \texttt{registerChanged} and \texttt{registerEvent} persists changes at the end of each service invocation, resulting in intermediate states that are visible before the entire functionality completes, leading to a lack of isolation. Therefore, it provides an API for the \textit{Application Layer} to manage these locks during the execution of system functionalities. Consequently, the commit implementation only needs to release the semantic locks, whereas abort must additionally trigger the execution of compensating transactions.
     \item \textbf{\texttt{Causal Unit Of Work Service}}: Implements a versioning optimistic concurrency control mechanism. Consequently, calling \texttt{registerChanged} and \texttt{registerEvent} at the end of a service invocation does not immediately persist changes. Instead, persistence occurs during the \texttt{commit} phase, which assigns a new version number. The \texttt{aggregateLoad} function then utilizes these version numbers to retrieve an aggregate version consistent with a causal snapshot. Since changes remain unwritten until the commit, \textit{abort} simply discards them. Furthermore, the commit process requires a version-merging step to prevent lost update anomalies when functionalities concurrently update the same aggregates due to optimistic concurrency control.
     Because this transactional model requires strict sequential version increments to establish causal history, it cannot utilize the distributed Snowflake ID generator of the \texttt{IVersion Service}; instead, it must rely on centralized versioning.
 \end{itemize}

  \item \textbf{\texttt{Unit Of Work}}: This abstract class is managed by the \texttt{Unit Of Work Service} and represents the transaction context that is passed through the command chain of invocation of a functionality. It is extended by \texttt{Saga Unit Of Work} and \texttt{Causal Unit Of Work}, which respectively hold the specialized runtime state required by their protocols (e.g., the list of modified aggregates, previous semantic states for compensation, or specific causal snapshots).

  \item \textbf{Aggregate Interfaces}: To actively participate in these distributed transactions, \textit{Application Layer} aggregates must implement these marker interfaces. \texttt{Saga Aggregate} forces the domain entity to hold a \texttt{Saga State} for semantic locking, while \texttt{Causal Aggregate} mandates the implementation of a \texttt{merge()} function to automatically resolve optimistic concurrency conflicts associated with the lost update anomaly.

  \item \textbf{Generic and Specific Commands}: Generic \texttt{Saga Command} and \texttt{Causal Command} abstract classes are provided to contain the payload required by the transactional model, like the semantic locks in the case of sagas. Specific commands, such as \texttt{Commit Saga Command} and \texttt{Commit Causal Command}, are defined for the infrastructural operations associated with transactional coordination.
 
  \item \textbf{Protocol-Specific Handlers (\texttt{Saga Command Handler} \& \texttt{Causal Command Handler})}: These classes explicitly implement the \texttt{Command Handler Decorator} interface exposed by the \textit{Messaging Module}. By acting as middleware decorators, they strategically intercept incoming \texttt{Command}s before they reach the pure domain logic. This mechanism is utilized to seamlessly process infrastructural operations (e.g., executing \texttt{Commit}, \texttt{Abort}, or \texttt{Prepare} commands) and to enforce transaction-specific invariants, such as verifying and acquiring semantic locks in Sagas. This ensures that the complex mechanics of distributed transaction coordination remain entirely abstracted from the application's business rules.

 \item \textbf{Coordination Extensions}: The \texttt{Workflow} class is extended to accommodate execution plans that vary by transactional model. For instance, for Sagas, an abort necessitates the execution of compensating transactions. Additionally, \textit{SagaStep} augments the standard execution logic by registering the required compensating transaction.
 
\end{itemize}

\section{Application Layer: The Quizzes System}
\label{sec:application_layer_quizzes}

While the \textit{Simulator Library} provides the foundation for coordination, messaging, and transactions, the \textit{Application Layer} is where the microservice systems are defined, containing their specific business logic and domain models. To demonstrate the simulator's capability to host complex distributed systems, this work utilizes a \textit{Quizzes} application as a comprehensive Domain-Driven Design (DDD) case study.

The \textit{Quizzes} application is structured into several aggregates, such as \textit{User}, \textit{Course}, \textit{Quiz}, and \textit{Tournament}, each managing its own state and invariants. By utilizing the \textit{Quizzes} application as a reference, the following subsections illustrate how the \textit{Application Layer}'s core modules, \textit{Application Domain} and \textit{Application Functionality}, are implemented by using and extending the \textit{Simulator Library}'s abstractions. This integration allows the \textit{Quizzes} system to remain entirely focused on its domain rules, while the simulator handles the complexities of data consistency across both centralized and distributed execution environments.

\subsection{Application Domain Module}
\label{subsec:application_domain}

\begin{figure*}[htbp]
    \centering
    \includegraphics[angle=90, height=0.8\textheight]{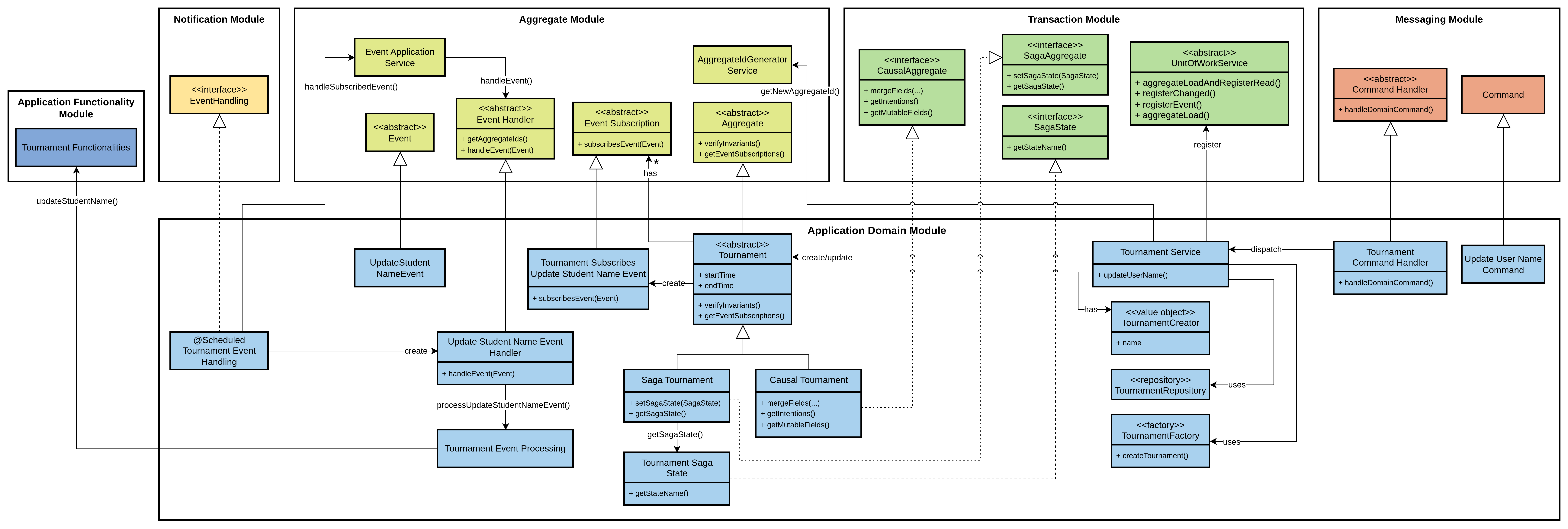}
    \caption{The architecture of the \textit{Application Domain Module} illustrates the \textit{Tournament} aggregate, demonstrating how Domain-Driven Design (DDD) concepts, such as \textit{Entities}, \textit{Value Objects}, \textit{Factories}, and \textit{Services}, operate alongside \textit{Upstream-Downstream} relationships within the simulator. Furthermore, this module details the implementation of the two transactional models within the microservice architecture.}
    \label{fig:simulator_layers_app_domain}
\end{figure*}

In the context of the \textit{Quizzes} application, the \textit{Application Domain Module} defines the clusters of domain objects that are treated as single units for data changes. Figure~\ref{fig:simulator_layers_app_domain} illustrates this architecture using the \texttt{Tournament} aggregate as a concrete example.

Every aggregate in the application extends the library's \texttt{Aggregate} abstract class. The \texttt{Tournament} entity maintains aggregate consistency by overriding the \texttt{verifyInvariants()} method; this ensures business rules, such as validating that \texttt{startTime} precedes \texttt{endTime}, are satisfied before state changes are committed. It includes the \texttt{Tournament Creator} \textit{Value Object}, which corresponds to a student from the \texttt{Course} aggregate. Additionally, \texttt{Tournament Service} defines the functional interface for the aggregate, while \texttt{Tournament Factory} handles the instantiation of specific \texttt{Tournament} instances according to the active transactional model.

For the realization of the upstream-downstream relationships, the \textit{Notification} and \textit{Messaging} modules are extended. 

The \texttt{Course} aggregate is upstream of the \texttt{Tournament}, and so, when a student name is changed, a \texttt{Update Student Name Event} is emitted. Figure~\ref{fig:simulator_layers_app_domain} traces the lifecycle of an \texttt{Update Student Name Event}. At the conceptual model level, the \texttt{Tournament} aggregate formally declares its reactive dependency on this event via its \texttt{getEventSubscriptions()} method (specifically returning a \texttt{Tournament Subscribes Update Student Name Event} contract). At runtime, the event processing cycle is initiated by the \texttt{Tournament Event Handling} component. Operating on a scheduled loop (\texttt{@Scheduled}) as dictated by the \textit{Notification Module}'s \texttt{Event Handling} interface, this component invokes the \texttt{handleSubscribedEvent} method of \texttt{Event Application Service}, passing both the target event type and its corresponding handler (\texttt{Update Student Name Event Handler}). 

Crucially, the \texttt{Event Application Service} and \texttt{Event Handler} reside within the \textit{Aggregate Module}. They assume responsibility for resolving subscriber aggregates and dynamically retrieving pending, matching events by querying the underlying \textit{Notification} infrastructure via the method \texttt{getSubscribedEvents} of the \texttt{Event Service}. Once retrieved, each matched event is dispatched to the concrete handler, which invokes the necessary domain processing logic (e.g., updating a student's name). This reaction ultimately triggers a new business process within the \textit{Tournament} functionalities.

To support direct communication, the \textit{Messaging} module is extended to define commands corresponding to each service in the \texttt{Tournament Service}, such as the \texttt{Update User Name Command}. Other aggregates use these commands to request actions from the \texttt{Tournament} aggregate. On the receiving side, a \texttt{Tournament Command Handler} is implemented to dispatch these commands to the appropriate service of the aggregate.

To execute according to a transactional model, the application aggregates must implement protocol-specific interfaces provided by the \textit{Transaction Module}:
\begin{itemize}
    \item \textbf{Saga Implementations (e.g., \texttt{Saga Tournament})}: Implement the \texttt{Saga Aggregate} interface, requiring the aggregate to manage a \texttt{Saga State}. This state acts as a semantic lock, allowing the system to track the aggregate's participation in long-running transactions and prevent conflicting operations.
    \item \textbf{Causal Implementations (e.g., \texttt{Causal Tournament})}: Implement the \texttt{Causal Aggregate} interface. This mandates the implementation of a \texttt{mergeFields()} method, enabling the optimistic concurrency control mechanism to resolve conflicts between concurrent versions of the aggregate to address the lost update anomaly.
\end{itemize}

The simulator decouples the tournament business logic from the specific transactional model in use. By defining \texttt{Tournament} as an abstract class with distinct implementations for each transactional model, the simulator facilitates the evaluation of different models (RQ2) while minimizing development effort (RQ4). Furthermore, because this approach only requires implementing the Domain-Driven Design (DDD) aspects of the domain (RQ1), it further addresses the goal of reducing overall complexity (RQ4). The same applies to the components that implement the remaining business logic, e.g. \texttt{TournamentService}, which are transactional model agnostic.

\subsection{Application Functionality Module}
\label{subsec:application_functionality}

\begin{figure*}[t]
    \centering
    \includegraphics[width=0.8\textwidth]{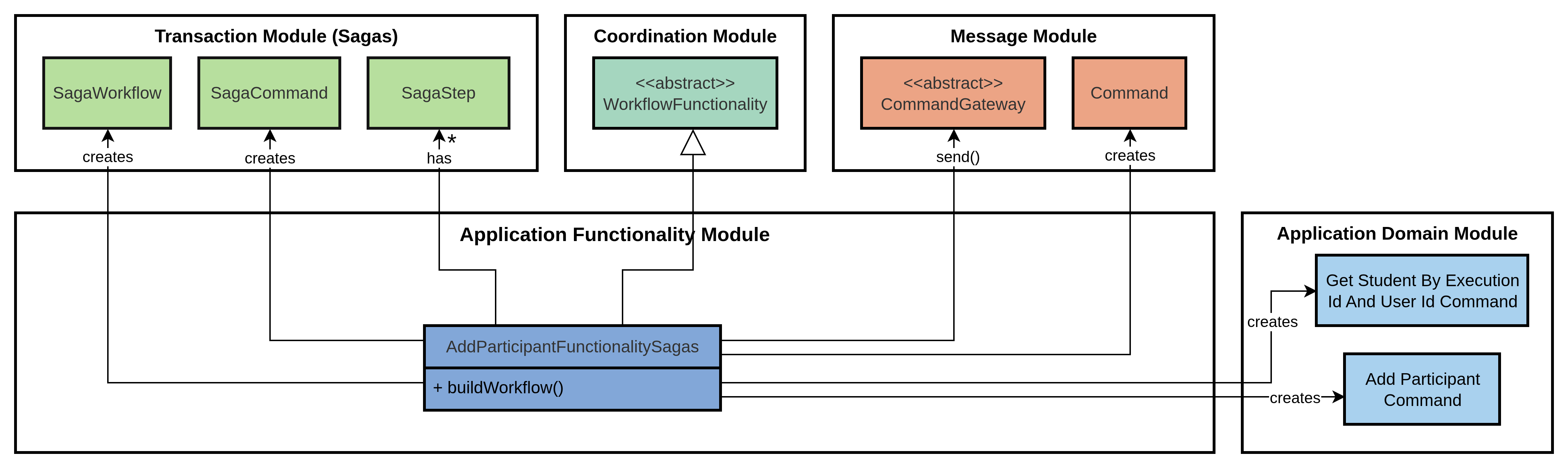}
    \caption{Architecture of the \textit{Application Functionality Module}, illustrating how a specific business process (e.g., \texttt{Add Participant Functionality Sagas}) extends the \texttt{Workflow Functionality} base class to orchestrate domain logic using steps and remote commands.}
    \label{fig:simulator_layers_app_functionalities}
\end{figure*}

The \textit{Application Functionality Module} encapsulates the high-level business processes and orchestration logic of the \textit{Quizzes} system. These functionalities represent the distinct use cases triggered by end-users or external systems. Figure~\ref{fig:simulator_layers_app_functionalities} shows how these processes are modeled, using the \texttt{Add Participant Functionality Sagas} as an illustrative example.

Rather than executing monolithic blocks of code, functionalities are implemented as structured workflows. By extending the \texttt{Workflow Functionality} class provided by the \textit{Coordination Module}, each \textit{Quizzes} functionality is required to define its execution sequence (e.g., via a \texttt{buildWorkflow()} method).

As depicted in the diagram, a concrete functionality interacts with several infrastructural components to orchestrate its logic for the Sagas transactional model:
\begin{itemize}
    \item \textbf{Workflow and Steps}: The functionality instantiates a protocol-specific workflow (e.g., a \texttt{Saga Workflow}) and populates it with discrete execution units (e.g., \texttt{Saga Step}s). This decomposition into steps is vital, as it enables the simulator to provide granular observability (tracing each step) and deterministic fault injection (failing or delaying specific steps).
    \item \textbf{Remote Commands}: To interact with other aggregates or to invoke actions on remote aggregates, the functionality use protocol-specific commands (e.g., a \texttt{Add Participant Command}).
    \item \textbf{Command Dispatching}: The functionality leverages the \textit{Messaging Module} by sending these constructed commands through the abstract \texttt{Command Gateway}. This abstracts the underlying network topology, allowing the functionality to invoke remote domain logic asynchronously or synchronously without managing the transport protocols.
\end{itemize}

\begin{lstlisting}[
		float=*t,
		language=Java, 
		caption={The \texttt{buildWorkflow()} of the \texttt{Add Participant Functionality Sagas}},
		label={lst:buildWorflow},
	    frame=single,
	    breaklines=true]
public void buildWorkflow(Integer tournamentAggregateId, Integer executionAggregateId, Integer userAggregateId,
            SagaUnitOfWork unitOfWork) {
    this.workflow = new SagaWorkflow(this, unitOfWorkService, unitOfWork);

    SagaStep getUserStep = new SagaStep("getUserStep", () -> {
        GetStudentByExecutionIdAndUserIdCommand getStudentCommand =
          new GetStudentByExecutionIdAndUserIdCommand(unitOfWork, ServiceMapping.EXECUTION.getServiceName(), ...);
        this.userDto = (UserDto) commandGateway.send(getStudentCommand);
    });

    SagaStep addParticipantStep = new SagaStep("addParticipantStep", () -> {
        List<SagaAggregate.SagaState> states = new ArrayList<>();
        states.add(TournamentSagaState.IN_UPDATE_TOURNAMENT);
        AddParticipantCommand addParticipantCommand
                    = new AddParticipantCommand(unitOfWork, ServiceMapping.TOURNAMENT.getServiceName(), ...);
        SagaCommand sagaCommand = new SagaCommand(addParticipantCommand);
        sagaCommand.setForbiddenStates(states);
        commandGateway.send(sagaCommand);
    }, new ArrayList<>(Arrays.asList(getUserStep)));

    this.workflow.addStep(getUserStep);
    this.workflow.addStep(addParticipantStep);
}
\end{lstlisting}
	
Listing~\ref{lst:buildWorflow} presents the detailed implementation of the \texttt{buildWorkflow} method of the \texttt{Add Participant Functionality Sagas}. This excerpt illustrates how the \textit{Application Developer} constructs a distributed transaction within the simulator. First, a protocol-specific \texttt{SagaWorkflow} is instantiated. Then, individual \texttt{SagaStep} objects are created to encapsulate remote interactions. For instance, the \texttt{addParticipantStep} instantiates a \texttt{AddParticipantCommand} that encapsulates the domain payload. Crucially, to enforce the Saga transactional model, the developer wraps this payload in a \texttt{SagaCommand} and defines semantic locks, establishing \texttt{forbiddenStates} to prevent the command from executing if the target tournament is currently locked in an \texttt{IN\_UPDATE\_TOURNAMENT} state. Furthermore, this step declares an explicit dependency on \texttt{getUserStep}, ensuring sequential execution. Finally, the \texttt{SagaCommand} is dispatched via the \texttt{Command Gateway}, and the steps are added to the workflow for the simulator to orchestrate.

Constructing a workflow for the Transactional Causal Consistency (TCC) model follows a nearly identical architectural pattern. Instead of configuring semantic locks, the developer wraps the domain payload in a \texttt{CausalCommand}. The overarching workflow and step dependencies remain structurally similar, but the burden of managing concurrent version conflicts is delegated entirely to the underlying \texttt{Causal Unit Of Work Service}, allowing the developer to rely on the model's snapshot isolation.

By relying on this architecture, the \textit{Quizzes} application maintains clean, cohesive business logic. When a user triggers an action like \textit{Add Participant}, the application simply defines the steps and commands required. The underlying \textit{Simulator Library} then assumes responsibility for orchestrating the execution, managing the transaction boundaries, and enforcing the selected consistency model.

The simulator requires distinct implementations for the same functionality to account for variations in transactional models and communication patterns (synchronous vs. asynchronous). This overhead is justified because the underlying semantics shift depending on the model; for instance, in Sagas, the use of semantic locks depends on the specific operation and its potential interactions with concurrent functionalities. Similarly, asynchronous communication allows for parallel command execution based on data dependencies, which significantly influences the overall design. For example, an asynchronous model typically returns an immediate acknowledgment to the caller while processing the task in the background.

\section{Deployment Topologies}
\label{sec:deployment_topologies}

The \textit{Microservice Simulator} is designed to be highly adaptable, allowing developers to execute the same core domain logic under entirely different deployment configurations. This capability is fundamental to the simulator's shift-left strategy. By supporting multiple deployment topologies, ranging from centralized, single-process executions to fully distributed microservice architectures, developers can observe how the business logic executing under different transaction models (e.g., Sagas, TCC) performs under varying network conditions and deployment models early in the design phase. 

Crucially, transitioning between these diverse deployment topologies requires zero code modifications or effort from the application developer. The infrastructural shift is completely transparent to the \textit{Application Layer} and is managed entirely through the application's environment configuration files.

\subsection{Centralized Local Deployment}
\label{subsec:topology_local}

\begin{figure*}[t]
    \centering
    \includegraphics[width=0.3\textwidth]{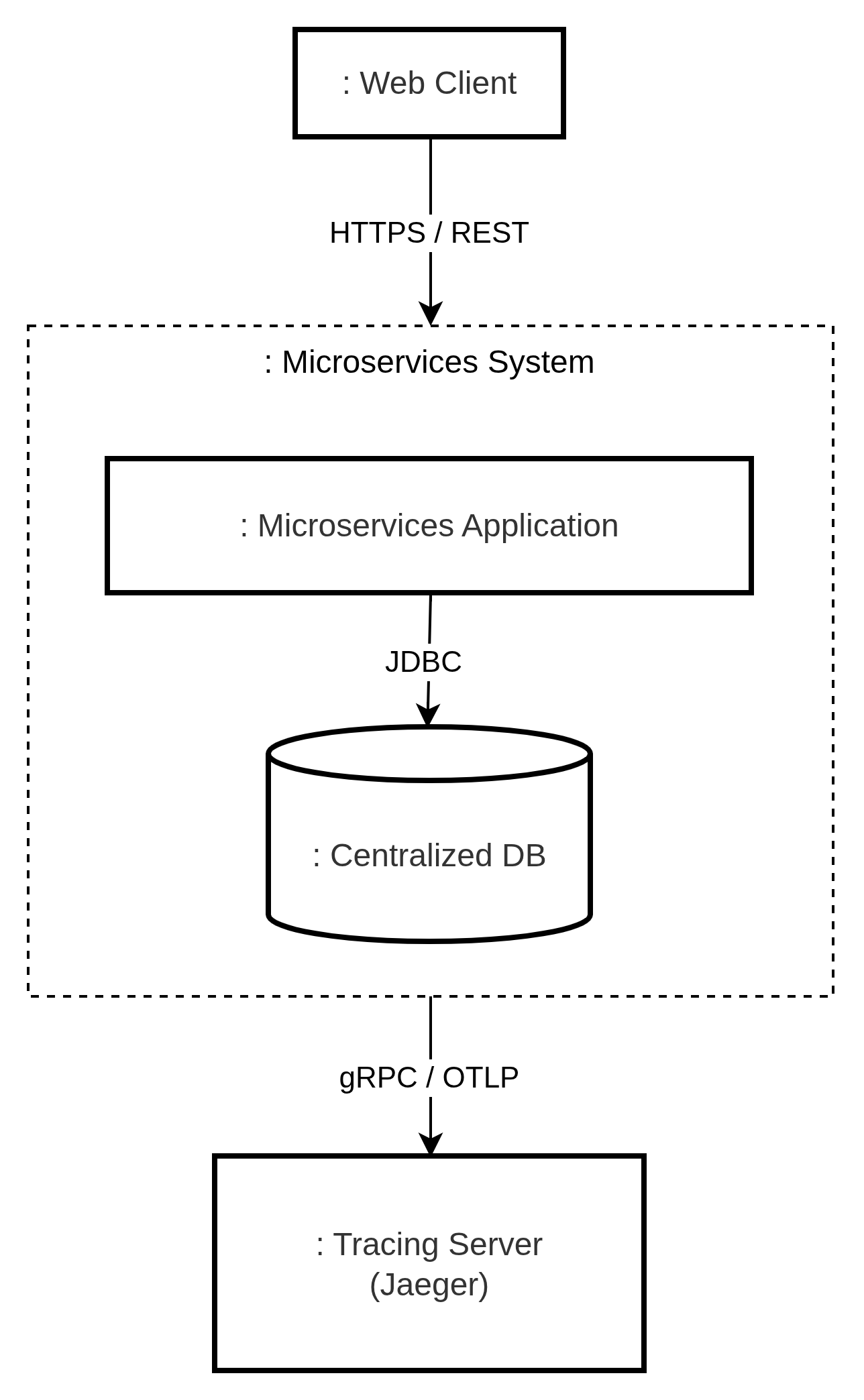}
    \caption{Architecture of the \textit{Centralized Local Deployment} topology, depicting a centralized execution environment where all aggregates communicate in-memory and share a single centralized database.}
    \label{fig:topology_local}
\end{figure*}

The simpler deployment model supported by the simulator is the \textit{Centralized Local Deployment}, illustrated in Figure~\ref{fig:topology_local}. In this configuration, the entire microservice system is executed as a single process \textit{:Microservices Application}, where all the modules execute. It facilitates the evaluation of microservices architectures within an economical and reproducible environment.

This topology is characterized by the following infrastructural choices:
\begin{itemize}
    \item \textbf{In-Memory Communication}: Instead of serializing commands over a network, inter-service communication and command routing occur via direct, synchronous method calls within the same Java Virtual Machine (JVM). Similarly, asynchronous event propagation bypasses external message brokers; aggregates publish domain events directly to a unified \texttt{Event Repository} hosted in the shared database. The scheduled polling loops of subscriber aggregates query this repository to fetch and process new events, fully preserving the decoupled, eventual consistency semantics without the overhead of network transport.
    \item \textbf{Centralized Database}: All application state is persisted to a single, shared relational database. However, the aggregate schemas remain logically decoupled. Interactions with the database are managed via standard JDBC/JPA connections.
    \item \textbf{Centralized Versioning}: The \texttt{Centralized Version Service} is employed, it executes in the main process where all other modules also execute (\texttt{:Microservices Application}), and utilizes the shared database (\texttt{:Centralized DB}) to maintain and increment version numbers sequentially.
\end{itemize}

The strategic value of this local deployment lies in providing a fast, deterministic baseline for a shift-left strategy. Developers can use this topology to seamlessly debug workflows by verifying business logic, simulating interleavings between functionalities in a deterministic context, and injecting faults or delays. Because execution is contained within a single JVM, developers can easily step through complex distributed transactions to ensure that aggregate boundaries, state transitions, and core business rules are logically sound.

Furthermore, despite the single-process execution, the simulator's observability mechanisms remain fully active. As shown in the diagram, the application continues to export distributed traces (via gRPC/OTLP) to an external tracing server (e.g., Jaeger). This allows developers to analyze the logically distributed workflows and transaction boundaries exactly as they would appear in a production environment. Ultimately, this makes the centralized local topology an ideal, noise-free sandbox for initial testing and validation before swapping configuration files to test the system in a complex, networked architecture.

\subsection{Initialization-Time Versioning Alternatives}
\label{subsec:topology_versioning}

A critical architectural choice in distributed systems is how to generate the deterministic version IDs required for concurrency control (particularly for protocols like TCC). For remote deployment topologies (such as Stream or gRPC), the simulator exposes an initialization-time alternative that drastically alters the network profile of the simulation:
\begin{enumerate}
    \item \textbf{Centralized System (Standalone)}: The simulator is configured to utilize a completely independent \textit{:Versioning Application} backed by a dedicated \textit{:Version DB}. To acquire a version ID, the main application must publish a request (e.g., via the Message Broker or a gRPC call) and await a response. This simulates a strict, centralized sequence generator but introduces significant network round-trip latency into the critical path of every distributed transaction.
    \item \textbf{Distributed System (In-Memory Snowflake)}: Alternatively, the system can be configured to use the \texttt{Distributed Version Service}, such that each microservice generates the version number locally, relying on the \texttt{Snowflake ID} algorithm~\cite{snowflake} to generate highly concurrent, collision-free identifiers entirely in-memory. This decentralized approach eliminates the network dependency on external services for ID generation, drastically reducing network overhead and simulating a highly scalable, independent node architecture.
\end{enumerate}

However, it is critical to note that this decentralized Snowflake optimization is incompatible with the \textit{Transactional Causal Consistency} (TCC) model because, in the current implementation, it is not possible to identify if the two coordinations are concurrent. A more complex distributed implementation is required.  Therefore, currently, as detailed in Section~\ref{subsec:transaction_module}, the TCC protocol's conflict resolution mechanics strictly mandate a centralized, monotonically increasing versioning system to guarantee strict sequence correctness.

\subsection{Centralized Stream Deployment}
\label{subsec:topology_stream}

\begin{figure*}[t]
    \centering
    \includegraphics[width=0.8\textwidth]{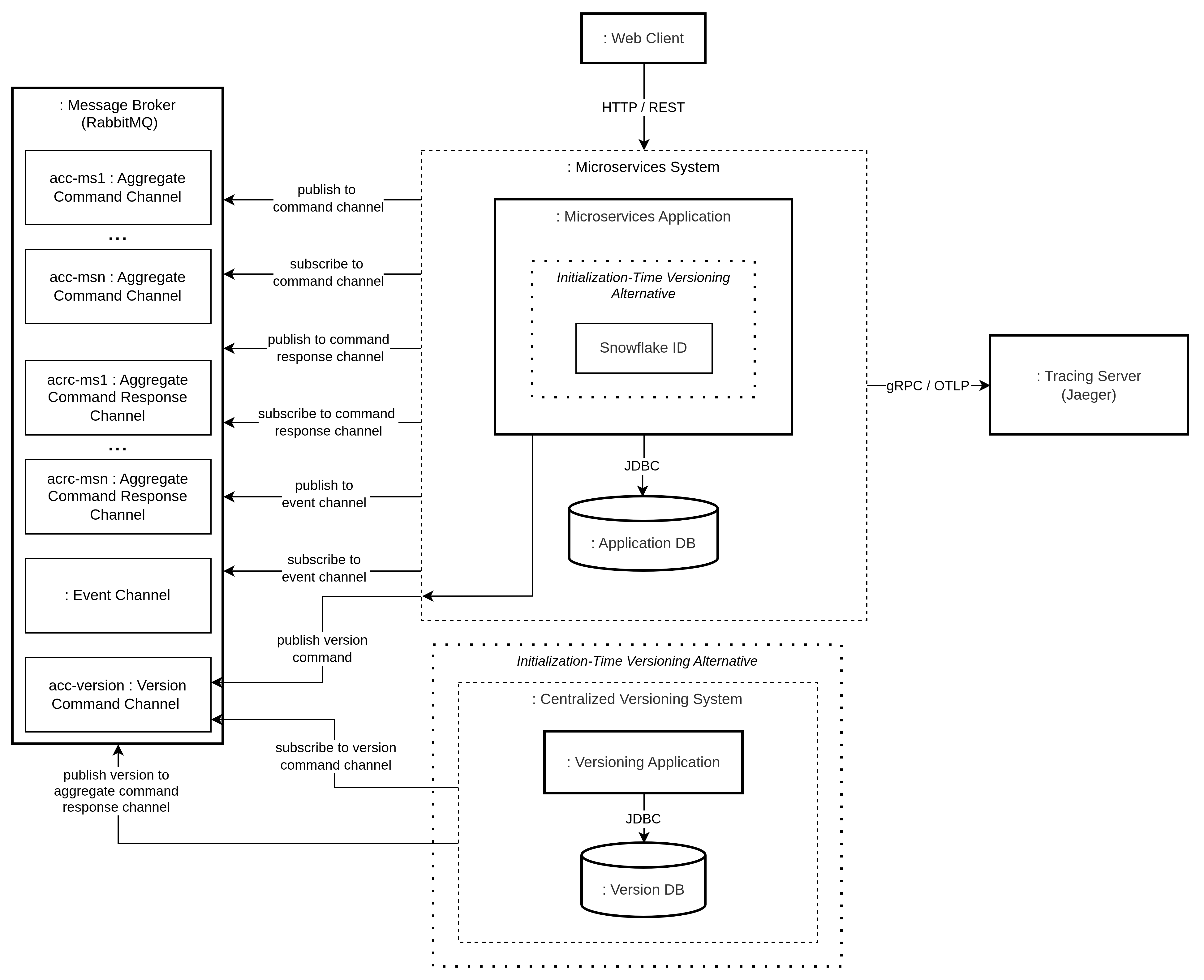}
    \caption{Architecture of the \textit{Centralized Stream Deployment} topology, introducing a \textit{Message Broker} (RabbitMQ) for inter-service communication while maintaining a shared, centralized application database. It also highlights the initialization-time alternative between a standalone database-backed \textit{Versioning Application} and an in-memory decentralized \texttt{Snowflake ID} generator.}
    \label{fig:topology_stream}
\end{figure*}

By altering the application's configuration profiles, developers can seamlessly transition the system to the \textit{Centralized Stream Deployment}, presented in Figure~\ref{fig:topology_stream}. This topology introduces network boundaries and messaging overhead into the execution environment while still relying on a centralized install of all modules in a single process (\texttt{:Microservice Application}) and a shared relational database (\texttt{:Application DB}) for domain state persistence. This offers the  advantage of enforcing a modular monolith~\cite{Su24} structure, where encapsulation between modules is enforced through remote communication.

The strategic value of this deployment model is that it acts as a first step between the local execution and a fully distributed microservices cluster~\cite{FAUSTINO2024102411}. It allows developers to test thoroughly and benchmark remote communication, such as the latency of distributed commands, the propagation of remote events, and the system's behavior, without the significant operational overhead of provisioning, managing, and debugging multiple independent application processes and individual databases.

In this configuration, all inter-context command communication is marshaled through a \texttt{:Message Broker} (RabbitMQ) via a publish/subscribe paradigm. To facilitate this messaging architecture, the simulator relies on specific logical channels mapped within the application's configuration files (e.g., \texttt{application.yaml}):
\begin{itemize}
    \item \textbf{Aggregate Command Channels (\texttt{acc-msn})}: Dedicated channels route commands to specific aggregates (e.g., \texttt{tournament-command-channel}). This ensures that workflows dispatch tasks strictly to the designated handler queues.
    \item \textbf{Aggregate Command Response Channels (\texttt{acrc-msn})}: Due to the decoupled nature of the broker, each aggregate maintains a dedicated response channel (e.g., \texttt{tournament-command-response-channel}). Remote services publish the outcome of their executions back to the specific aggregate response channel, allowing the orchestrating workflow to retrieve the result and proceed.
    \item \textbf{Event Channel}: A separate, unified pub/sub channel (\texttt{event-channel}) is dedicated exclusively to broadcasting domain events. The \texttt{Event Publisher Service} periodically polls the local database for unpublished events and dispatches them to this channel. Concurrently, the \texttt{Event Subscriber Service} listens to this channel, filtering and persisting relevant events into the local repository of the subscribing microservice for subsequent processing.
    \item \textbf{Version Command Channel (\texttt{acc-version})}: As detailed in Section~\ref{subsec:topology_versioning}, if the standalone \texttt{:Versioning Application} is deployed, microservices use this dedicated channel to request version IDs for concurrency control. For instance, the \texttt{:Microservices Application} publishes a request to the \texttt{acc-version} channel and blocks, waiting to receive the generated version ID via its own dedicated response channel (\texttt{acrc-msn}). The \textit{:Versioning Application} consumes the request from the command channel and routes the response strictly back to the caller's specific response queue. If the in-memory alternative is configured, this channel is bypassed entirely, and versioning is handled locally via the \texttt{:Distributed Version Service}.
\end{itemize}

\subsection{Centralized gRPC Deployment}
\label{subsec:topology_grpc}

\begin{figure*}[t]
    \centering
    \includegraphics[width=0.8\textwidth]{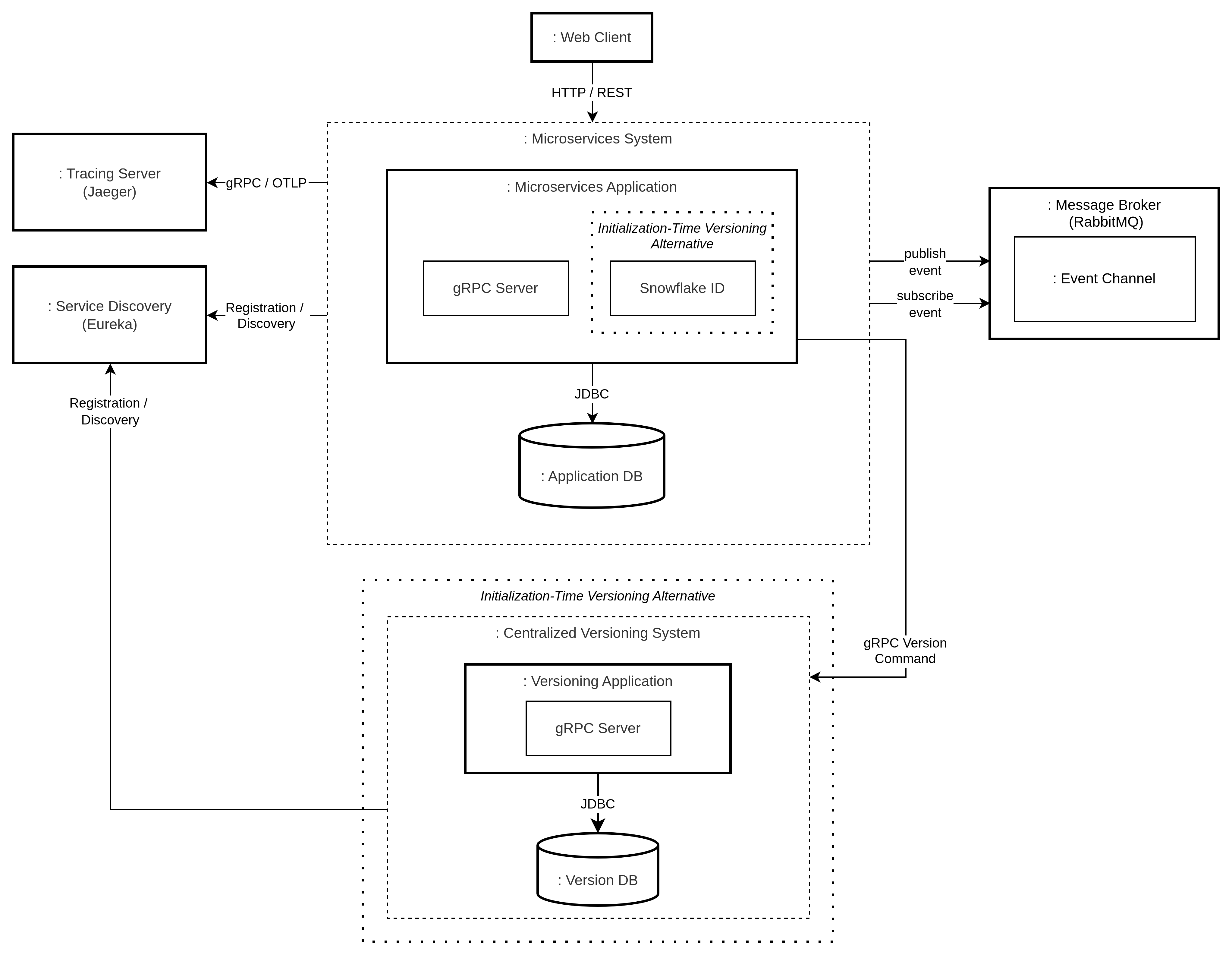}
    \caption{Architecture of the \textit{Centralized gRPC Deployment} topology. It utilizes a gRPC server for point-to-point command communication, while retaining RabbitMQ exclusively for event broadcasting. The shared \texttt{:Application DB} and versioning alternatives remain identical to the \textit{Stream} topology. Additionally, it incorporates \texttt{:Service Discovery} server (e.g., Eureka) to register and resolve network locations, including interactions with the standalone \textit{:Versioning Application}.}
    \label{fig:topology_grpc}
\end{figure*}

The \textit{Centralized gRPC Deployment}, illustrated in Figure~\ref{fig:topology_grpc}, provides an alternative network architecture for developers seeking to evaluate point-to-point communication. Much like the \textit{Stream} topology, its primary strategic value lies in acting as a controlled intermediate state. It allows developers to thoroughly test and benchmark remote procedure calls, service discovery resolution, and network latency without the operational burden of managing a fully distributed database cluster. This represents an alternative implementation of modular monolith enforcement~\cite{Su24}, supported by the \textit{Microservice Simulator}

In this configuration, the underlying communication mechanisms are fundamentally altered:
\begin{itemize}
    \item \textbf{Commands via gRPC}: Inter-service commands are dispatched by the \texttt{Grpc Command Gateway} as Remote Procedure Calls (RPCs) supported by a dedicated internal \texttt{gRPC Server} library running in the \texttt{:Microservices Application}. 
    \item \textbf{Service Discovery}: To route calls dynamically, the topology introduces a \texttt{:Service Discovery} registry (e.g., Eureka). When a workflow dispatches a command, the system queries the registry to resolve the network location of the target aggregate before initiating the gRPC connection. In this centralized configuration, all inter-aggregate commands naturally resolve to the same server (\texttt{:Microservices Application}). However, relying on the registry ensures that the command routing logic remains entirely identical to a fully distributed setup, while also enabling the dynamic discovery of the external, standalone \textit{:Versioning Application}.
    \item \textbf{Event Channel (RabbitMQ)}: Crucially, while commands become direct RPCs, the architecture retains the\texttt{:Message Broker} (RabbitMQ) strictly for the \texttt{:Event Channel}. This hybrid approach ensures that domain events remain fully decoupled, allowing the system to react to state changes without blocking primary execution threads. The implementation is like in the \textit{Stream} topology.
    \item \textbf{Version Commands via gRPC}: As introduced in Section~\ref{subsec:topology_versioning}, if the standalone \textit{:Versioning Application} is deployed, it is queried via dedicated gRPC version commands, such as \texttt{Get Version Command}, adding RPC network hops for concurrency ID generation. If the \texttt{distributed-version} profile is active, these remote calls are bypassed entirely in favor of the local, in-memory \textit{Snowflake} generator.
\end{itemize}

\subsection{Distributed Stream Deployment}
\label{subsec:topology_distr_stream}

\begin{figure*}[t]
    \centering
    \includegraphics[width=1\textwidth]{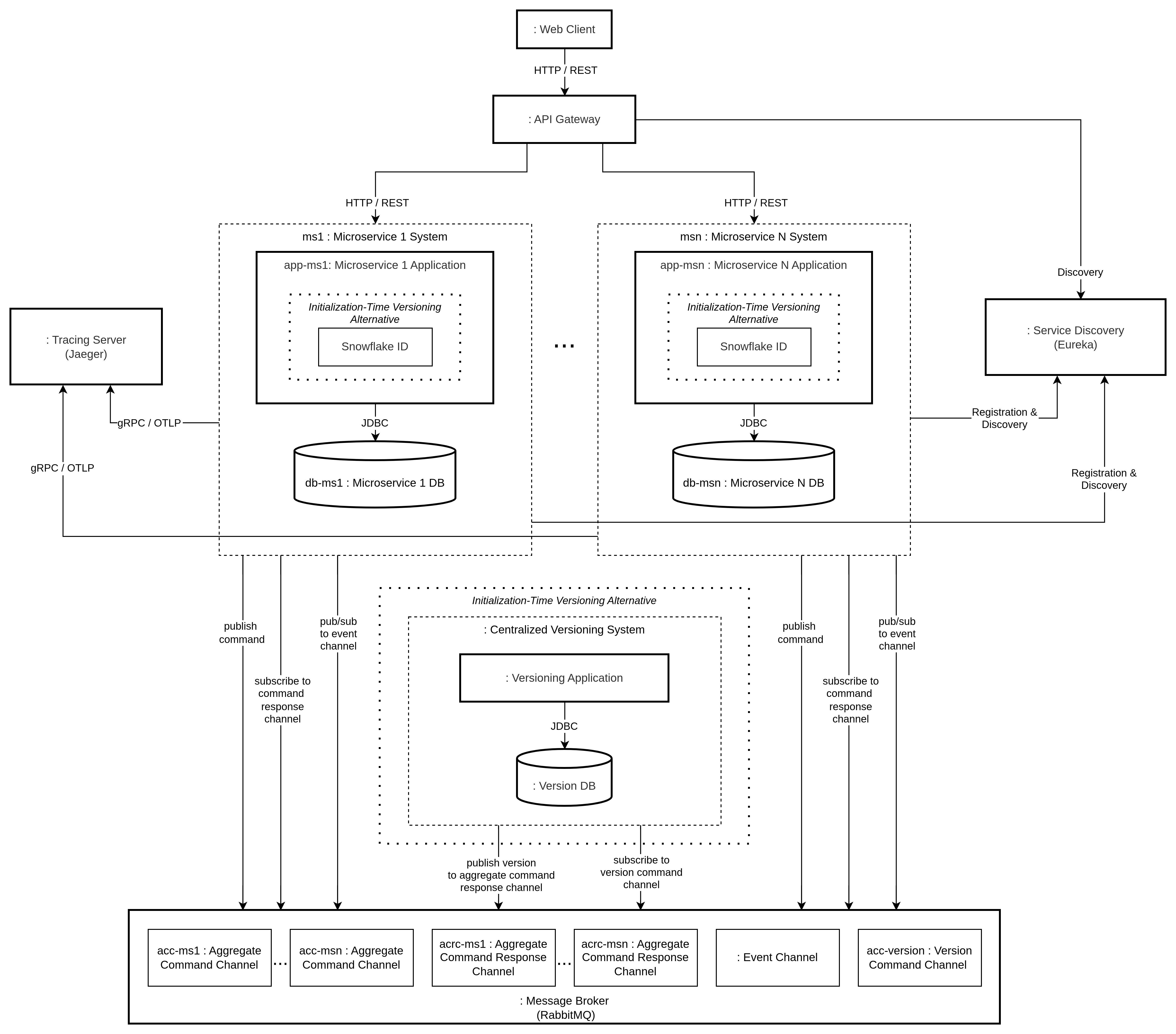}
    \caption{Architecture of the \textit{Distributed Stream Deployment} topology. This fully decentralized configuration features independent microservices (e.g., ms1, ms2) with isolated databases, communicating entirely via a \texttt{:Message Broker} (RabbitMQ) queues. \texttt{:Service Discovery} (Eureka) is utilized for initial routing, while traces are aggregated centrally via Jaeger.}
    \label{fig:topology_distr_stream}
\end{figure*}

The \textit{Distributed Stream Deployment}, depicted in Figure~\ref{fig:topology_distr_stream}, represents the transition from a centralized environment to a fully decentralized, production-grade microservices architecture. In this topology, the unified application runtime and shared database are completely redesigned. 

Instead, the domain logic is partitioned into strictly isolated, independent microservice processes (e.g., \texttt{app-ms1}, \texttt{app-ms2}). Crucially, each microservice assumes exclusive ownership of its persistence layer, interacting only with its dedicated database (e.g., \texttt{db-ms1} to \texttt{db-msn}) via standard JDBC connections. This strict data segregation adheres to the microservices principle of database-per-service, ensuring each component maintains its own isolated data store.

To manage external client interactions across the independent microservice processes, a central entry point is required. To address this, the simulator provides a dynamic \texttt{:API Gateway}. Unlike traditional gateways that require static, centralized configuration, this component is designed for maximum reusability. It operates as an application utilizing a custom dynamic proxy controller. Rather than hardcoding network routes, the gateway leverages the \texttt{:Service Discovery} registry (e.g., Eureka) to resolve the active network locations of the deployed microservices. It then performs an HTTP \texttt{GET} request to retrieve the specific route mappings defined within each microservice's individual configuration file. This decentralized routing strategy ensures that as developers add new aggregates or modify their APIs, the gateway automatically adapts to forward REST requests to the correct application process, completely abstracting the network routing complexity away from the application developer.

To facilitate coordination across microservices, the architecture relies exclusively on a \texttt{:Message Broker} (RabbitMQ) implementing a publish/subscribe paradigm. Because the underlying messaging infrastructure is decoupled from the physical deployment, the routing of inter-service messages is governed by the same network of dedicated logical channels utilized in the \textit{Centralized Stream Deployment} (Section~\ref{subsec:topology_stream}). 

Each independent microservice exposes its own dedicated command queue (\texttt{acc-msn}) and response channel (\texttt{acrc-msn}), while continuing to utilize the unified \texttt{event-channel} for broadcasting domain events and the \texttt{acc-version} channel for centralized ID generation (unless the \texttt{distributed-version} profile is active).

To manage the initial entry points into this distributed network, the topology introduces a \texttt{:Service Discovery} registry (e.g., Eureka) and an \textit{:API Gateway}. The \texttt{:Web Client} interacts with the \texttt{:API Gateway}, which resolves the active network locations of the independent microservices to route external HTTP/REST requests appropriately. Despite the physical decoupling, the simulator maintains holistic observability by having every independent node export its telemetry data via gRPC/OTLP to a centralized \textit{:Tracing Server} (e.g., Jaeger), enabling researchers to visualize cross-node transaction flows.

\subsection{Distributed gRPC Deployment}
\label{subsec:topology_distr_grpc}

\begin{figure*}[t]
    \centering
    \includegraphics[width=1\textwidth]{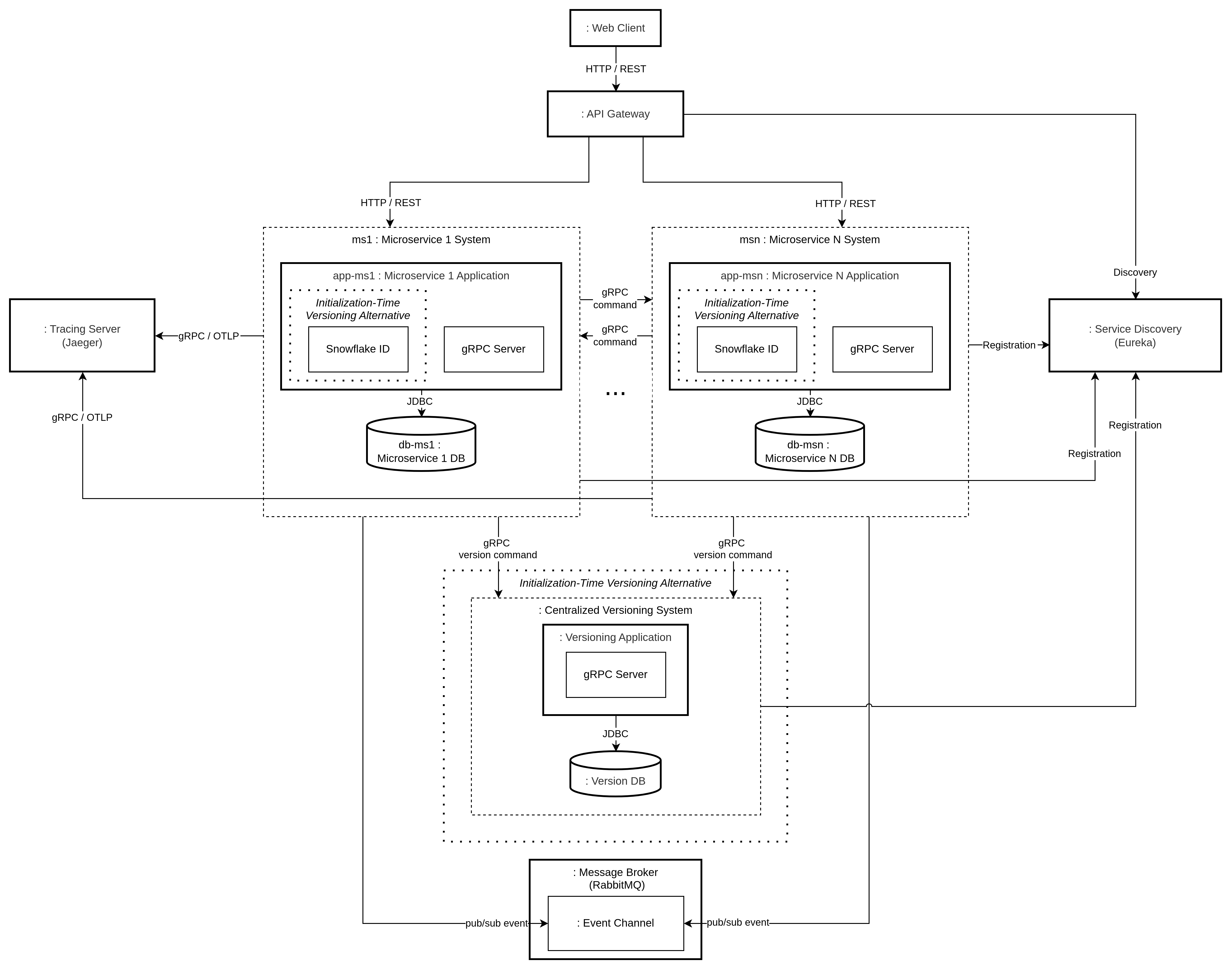}
    \caption{Architecture of the \textit{Distributed gRPC Deployment} topology. In this fully decentralized model, isolated microservices execute remote commands via point-to-point gRPC Remote Procedure Calls (RPCs) resolved through a \textit{:Service Discovery} (Eureka). A \textit{:Message Broker} (RabbitMQ) is retained exclusively for broadcasting domain events.}
    \label{fig:topology_distr_grpc}
\end{figure*}

The \textit{Distributed gRPC Deployment}, illustrated in Figure~\ref{fig:topology_distr_grpc}, represents an alternative production-grade, decentralized architecture. Similar to the \textit{Distributed Stream} topology, it redesigns the monolithic execution environment, partitioning the domain logic into strictly isolated microservice processes (e.g., \texttt{app-ms1}, \texttt{app-msn}), each possessing exclusive access to its own dedicated database (e.g., \texttt{db-ms1}, \texttt{db-msn}).

The defining characteristic of this topology is its hybrid approach to inter-service communication, purposefully mixing point-to-point routing with broadcasting to optimize specific coordination requirements. Because the communication layer is completely decoupled from the process boundaries, this deployment inherits the same communication mechanics as the \textit{Centralized gRPC Deployment} (Section~\ref{subsec:topology_grpc}).

Inter-service commands and centralized version requests are dispatched as direct \textit{Remote Procedure Calls} (RPCs), relying on the Service Discovery registry (e.g., Eureka) to dynamically resolve the target microservice's active network location. Simultaneously, the architecture retains the \texttt{:Message Broker} (RabbitMQ) strictly for broadcasting and subscribing to domain events, ensuring decoupled, reactive side effects without blocking the primary gRPC execution threads.

External interactions are managed via an \texttt{:API Gateway}, which also utilizes the \texttt{:Service Discovery} registry to route client requests to the appropriate entry-point microservice. Furthermore, cross-node observability is maintained by having every independent service export telemetry data via gRPC/OTLP to a centralized Tracing Server (e.g., Jaeger).

\subsection{Containerized Deployments}
\label{subsec:topology_docker_k8s}

To bridge the gap between local JVM execution and production-grade operational environments, the \textit{Microservice Simulator} supports containerized deployments. Crucially, to minimize the effort required by developers to transition between the deployment topologies described previously, the simulator relies heavily on Spring Boot profiles and decoupled container orchestration scripts.

\subsubsection{Centralized vs. Distributed Execution Models}
The fundamental shift between the tested topologies dictates how the application is packaged and executed. Developers can seamlessly transition between these models by changing the application's entry point:
\begin{itemize}
    \item \textbf{Centralized Topologies}: To execute the \textit{Centralized Local}, \textit{Centralized Stream}, or \textit{Centralized gRPC} deployments, the developer builds a single Docker image representing the unified \texttt{:Microservices Application}. This image runs a single Spring Boot \texttt{main} method that instantiates all aggregates and modules simultaneously within one container process.
    \item \textbf{Distributed Topologies}: To execute the \textit{Distributed Stream} or \textit{Distributed gRPC} deployments, the centralized execution model is bypassed. The developer builds independent Docker images for each aggregate (e.g., \texttt{tournament-service}, \texttt{quiz-service}). Each container executes its own dedicated Spring Boot \texttt{main} method, guaranteeing strict process, memory, and database isolation.
\end{itemize}

\subsubsection{Docker Compose Execution}

For local distributed testing of these topologies, the simulator utilizes a two-layer \textit{Docker Compose} architecture to cleanly separate core infrastructure from domain-specific applications. This setup provides an accessible environment for validating inter-service communication before moving to more complex orchestrators:
\begin{itemize}
    \item \textbf{Root Configuration}: A common \texttt{docker-compose.yml} file at the repository root defines the shared infrastructural backbone. This includes the requisite middleware (\textit{RabbitMQ}), observability tools (\textit{Jaeger}), service discovery (\textit{Eureka}), and the \texttt{API Gateway}.
    \item \textbf{Application Configuration}: Each specific domain model (e.g., the \textit{Quizzes} application) maintains its own \texttt{docker-compose.yml} file that inherits and extends the root configuration. This application-level file defines the specific microservice containers and their dedicated databases. To instantiate a specific topology (e.g., \textit{Distributed Stream} vs. \textit{Distributed gRPC}), the developer does not need to rewrite container definitions; they simply pass the desired Spring profiles (e.g., \texttt{stream}, \texttt{grpc}, \texttt{sagas}, \texttt{tcc}) as environment variables when executing the compose command.
\end{itemize}

\subsubsection{Kubernetes Orchestration}

While Docker Compose excels at local integration testing, transitioning to \textit{Kubernetes} introduces the robust orchestration, self-healing, and dynamic configuration mechanisms characteristic of true production-grade environments. Moving beyond local containers, the simulator supports deployment to resilient, highly available Kubernetes clusters. 

To execute this transition for a domain application, the developer replaces the \texttt{docker-compose.yml} with a set of declarative Kubernetes manifests. The simulator provides the baseline infrastructure definitions (RabbitMQ, Jaeger, PostgreSQL), so the developer is only responsible for creating the following deployment artifacts for their specific microservices:

\begin{itemize}
    \item \textbf{Deployment Manifests}: For each isolated aggregate (e.g., \texttt{tournament-deployment.yaml}), the developer must define a Kubernetes \texttt{Deployment}. This specifies the container image, resource limits, and maps the required Spring Boot profiles as environment variables.
    \item \textbf{Service Manifests}: To replace Eureka, the developer must create a Kubernetes \texttt{Service} for each aggregate. This exposes the microservice internally within the cluster, allowing the \textit{Spring Cloud Kubernetes} library to seamlessly resolve network locations via native Kubernetes DNS (e.g., for point-to-point gRPC calls).
    \item \textbf{ConfigMaps and Secrets}: Instead of relying entirely on environment variables passed at startup, shared environment parameters (such as active messaging layers, transaction profiles, and database credentials) are fully abstracted from the container images. The developer defines these in a \texttt{ConfigMap}, which is dynamically injected into the pods at runtime, allowing configuration updates without rebuilding images.
    \item \textbf{RBAC Configurations}: Because the microservices must communicate directly with the Kubernetes API to resolve service locations, the deployment mandates strict security policies that were not required in local Docker networks. The developer must apply the provided Role-Based Access Control (RBAC) manifests, which create a \texttt{ServiceAccount} and \texttt{RoleBinding} to grant the application pods the precise, read-only permissions required to monitor endpoints within their designated namespace.
\end{itemize}

\subsubsection{Cloud Integration (Azure AKS)}

While the provided Kubernetes manifests can be executed on local clusters (e.g., Minikube or Kind), transitioning to a managed cloud environment provides the ultimate validation of scalability and network latency. The repository is equipped for straightforward cloud transitions. A dedicated deployment script (\texttt{push-to-acr.sh}) automates the cloud migration process. It provisions an Azure Container Registry (ACR), tags the locally built Docker images, and pushes them to the remote registry. The script subsequently binds the ACR to an Azure Kubernetes Service (AKS) cluster, establishing a secure pipeline. Once the images are hosted in the cloud, developers can apply the Azure-optimized Kubernetes manifests to deploy the supporting infrastructure (PostgreSQL, RabbitMQ, Jaeger) alongside the distributed microservice pods, achieving a fully managed, cloud-native simulation environment.

\section{Evaluation}
\label{sec:evaluation}

The evaluation of the \textit{Microservice Simulator} is structured to systematically address the research questions posed in Section~\ref{sec:introduction}. We first do an analysis of the simulator coverage of the DDD concepts to address RQ1. Then, also to address RQ1, we present the implementation of a complex case study, the \textit{Quizzes} system. Next, we evaluate the resilience and developer experience of diverse transactional models, specifically \textit{Sagas} and \textit{Transactional Causal Consistency} (RQ2). We then present comprehensive performance benchmarks across various centralized and distributed deployment topologies to show how the decoupling of business logic from communication infrastructure (RQ3) can facilitate performance benchmarking. Finally, we discuss the modularity of the simulator's architecture, analyzing the effort required for both application developers to implement their domain logic and library developers to extend the simulator with new models (RQ4).

\subsection{Expressiveness}
\label{sec:expressiveness}

An important aspect of the \textit{Microservice Simulator} is the extent to which it supports the diverse architectural patterns found in Domain-Driven Design (DDD) and distributed systems (RQ1). To provide a comprehensive foundation for shift-left validation and facilitate rigorous domain modeling, the simulator includes abstractions for foundational DDD patterns. Table~\ref{tab:ddd_completeness} summarizes these primary structures and details how they are currently supported within the simulator's architecture.

\begin{table}[h!]
\renewcommand{\arraystretch}{1.3}
\setlength{\tabcolsep}{8pt}
\centering
\caption{Support for Domain-Driven Design (DDD) patterns within the Microservice Simulator.}
\label{tab:ddd_completeness}
\resizebox{\columnwidth}{!}{
\begin{tabular}{|l|p{8cm}|}
\hline
\textbf{DDD Pattern} & \textbf{Simulator Support \& Implementation} \\ \hline
\textbf{Bounded Context} & \textbf{Partially Supported.} Mapped to independent, deployable microservices. However, the current evaluation enforces a strict one-to-one mapping between an aggregate and a microservice, simplifying the traditional modeling of multiple aggregates within a single bounded context. \\ \hline
\textbf{Aggregate} & \textbf{Fully Supported.} Implemented via the base \texttt{Aggregate} class, managing state evolution, versioning, and intra-aggregate consistency (invariants). \\ \hline
\textbf{Entity} & \textbf{Fully Supported.} Modeled as standard JPA entities within the aggregate boundary. \\ \hline
\textbf{Value Object} & \textbf{Fully Supported.} Modeled as immutable objects embedded within entities. \\ \hline
\textbf{Domain Event} & \textbf{Fully Supported.} Managed by the \textit{Notification Module} for asynchronous communication between bounded contexts. \\ \hline
\textbf{Command} & \textbf{Fully Supported.} Managed by the \textit{Messaging Module} for explicit, targeted remote invocations. \\ \hline
\textbf{Factory} & \textbf{Fully Supported.} Encapsulates complex aggregate instantiation logic, dynamically generating model-specific entities (e.g., \texttt{SagaTournament}). \\ \hline
\textbf{Service} & \textbf{Fully Supported.} Services are realized via \texttt{*Service} classes (e.g., \texttt{CourseService}) orchestrating the \texttt{Workflow Functionality} and transaction boundaries. Pure Domain Services are minimized in favor of a Rich Domain Model. \\ \hline
\textbf{Repository} & \textbf{Fully Supported.} Implemented via standard Spring Data JPA interfaces, maintaining strict database-per-service isolation. \\ \hline
\textbf{Anti-Corruption Layer} & \textbf{Implicitly Supported.} \texttt{Command}, \texttt{Event}, and \texttt{DTO} classes act as a stable published language over the wire, ensuring internal JPA \texttt{Entities} are never leaked or corrupted across service boundaries. \\ \hline
\textbf{Specification} & \textbf{Not Explicitly Used.} Complex business rules are enforced directly within the Aggregate's \texttt{verifyInvariants()} method rather than being encapsulated in standalone Specification objects. \\ \hline
\end{tabular}%
}
\end{table}

Additionally, the simulator provides an explicit support for microservices coordination through orchestrations. To achieve orchestration using choreography, inter-microservices communication occurs only through the \textit{Notification Module}, there are no upstream-downstream relationships, and there is maximum decoupling between microservices. Therefore, developers have to define functionalities that act strictly as reactive event processors. As verified in the implemented codebase (e.g., \texttt{Anonymize User Tournament Functionality Sagas}), instead of coordinating multiple remote services, the functionality's workflow consists of a single execution step that dispatches a command exclusively to its \textit{primary} aggregate service. Once the local service mutates the state, it publishes a new domain event. The \textit{Notification Module} then dynamically routes this event to subscribers, triggering their respective local workflows. Consequently, the \texttt{Workflow Functionality} is utilized solely to enforce the local transactional lifecycle. Because these single-step functionalities commit and release their locks immediately, the simulator cannot automatically do compensations. This completely shifts the burden of rollback onto the microservices themselves, requiring developers to manually design and emit explicit compensating events to reverse unlocked, committed states. This resulting increase in business logic complexity directly corresponds to the criticism of choreography coordination~\cite{richardson19}, which struggles to scale for complex, multi-step interactions between microservices.

\subsection{Experimental Setup: The Quizzes System}
\label{subsec:evaluation_quizzes}

To empirically evaluate the \textit{Microservice Simulator} and directly address RQ1 (facilitating DDD experimentation), a significant part of an existing business-logic-rich monolith system, the \textit{Quizzes}\footnote{\href{https://quizzes-tutor.tecnico.ulisboa.pt/}{https://quizzes-tutor.tecnico.ulisboa.pt/}} application, was decomposed into a set of aggregates. Rather than relying on trivial examples or synthetic benchmarks devoid of domain semantics, this adaptation provides a realistic baseline characterized by strict invariants, complex transactions, and intricate aggregate dependencies.

Our evaluation focuses on fully utilizing the distributed and decoupled capabilities of the simulator. The system's domain model was decomposed into eight distinct aggregates, each strictly isolated and managed by its own independent microservice: \textit{Course}, \textit{User}, \textit{Topic}, \textit{Question}, \textit{CourseExecution}, \textit{Quiz}, \textit{Answer}, and \textit{Tournament}. To quantify the domain complexity encapsulated within these boundaries, Table~\ref{tab:aggregate-metrics} details the number of internal business invariants enforced before any state mutation, the number of asynchronous event subscriptions (representing upstream dependencies), and the total number of exposed API methods per aggregate.

\begin{table}[h!]
\centering
\caption{Distribution of invariants, event subscriptions, and Commands (API) across the Quizzes system aggregates.}
\label{tab:aggregate-metrics}
\resizebox{\columnwidth}{!}{
\begin{tabular}{|l|c|c|c|}
\hline
\textbf{Aggregate}       & \textbf{Invariants} & \textbf{Event Subscriptions} & \textbf{Commands (API)} \\ \hline
Course                   & 1                   & 0                            & 7                       \\ \hline
User                     & 1                   & 0                            & 7                       \\ \hline
Topic                    & 0                   & 0                            & 5                       \\ \hline
Question                 & 1                   & 2                            & 9                       \\ \hline
CourseExecution          & 1                   & 1                            & 11                      \\ \hline
Quiz                     & 4                   & 3                            & 11                      \\ \hline
Answer                   & 0                   & 5                            & 9                       \\ \hline
Tournament               & 12                  & 8                            & 19                      \\ \hline
\end{tabular}%
}
\end{table}

The data reveals a distinct architectural topology inherent to DDD. Core, upstream aggregates like \textit{Course} possess no event subscriptions, as they act as the core modules. Conversely, downstream aggregates like \textit{Tournament} are significantly more complex. To maintain autonomy while reacting to changes in upstream services (e.g., a \textit{User} updating their name), the \textit{Tournament} aggregate manages 8 distinct event subscriptions and enforces 12 complex internal invariants. To support this in a truly distributed environment, the application utilizes the simulator's distributed event processing, enabling asynchronous event propagation across the network via a message broker while preserving the aggregate's pure domain logic.

Furthermore, to fully validate the simulator's decoupled coordination capabilities, the application's inter-aggregate interactions strictly avoid direct service invocations. Instead, the system uses explicit, network-agnostic \textit{Commands} routed through the simulator's \textit{CommandGateway}. This allows us to securely orchestrate a total of 45 distinct business functionalities across the isolated microservices. To understand the coordination complexity, we categorized these functionalities into four levels of transactional complexity. Each functionality is associated with an aggregate, e.g., \texttt{AddParticipantFunctionality} functionality and \texttt{Tournament} aggregate, its primary aggregate, and, consequently, it can invoke query and update services in all the upstream (secondary) aggregates, and the result of its execution can emit events to be consumed by downstream aggregates.

\begin{itemize}
    \item \textbf{Type A (Query)}: The functionality strictly performs distributed reads across the primary and/or secondary upstream aggregates without mutating state.
    \item \textbf{Type B (Simple Functionality)}: The functionality modifies the state of its primary aggregate exclusively, though it may require reads from remote upstream aggregates to obtain data and validate business rules before the mutation.
    \item \textbf{Type C (Complex Functionality)}: The functionality dictates a distributed transaction, explicitly writing to both its primary aggregate and one or more secondary aggregates via remote commands.
    \item \textbf{Type D (Event Functionality)}: The functionality mutates its primary aggregate and consequently publishes a domain event. This triggers asynchronous, eventual consistency workflows in secondary downstream aggregates that subscribe to that state change.
\end{itemize}

Note that these categories are not mutually exclusive; a single operation, such as updating a student's name, can behave as both a Type B functionality (updating the \textit{User} aggregate) and a Type D functionality (publishing an event that updates the name within the \textit{Tournament} aggregate). Table~\ref{tab:functionality-complexity} summarizes the distribution of these complexities across the implemented system.

\begin{table}[h!]
\centering
\caption{Complexity distribution of the 45 functionalities orchestrated within the Quizzes system.}
\label{tab:functionality-complexity}
\begin{tabular}{|c|c|}
\hline
\textbf{Complexity Type} & \textbf{Number of Functionalities} \\ \hline
Type A (Query)           & 17                                 \\ \hline
Type B (Simple)          & 24                                 \\ \hline
Type C (Complex)         & 4                                  \\ \hline
Type D (Event)           & 10                                 \\ \hline
\end{tabular}
\end{table}

To quantify the domain complexity encapsulated within these boundaries, Table~\ref{tab:ddd-structural-metrics} summarizes the system's structural DDD metrics. The \textit{Quizzes} case study contains 8 distinct aggregates and 8 aggregate services, which actively coordinate 45 functionalities. We enforce aggregate-level consistency through 20 total invariants (with the per-aggregate distribution detailed previously in Table~\ref{tab:aggregate-metrics}). Furthermore, we capture cross-aggregate consistency through 10 distinct upstream-downstream relations. As detailed in Table~\ref{tab:upstream-downstream-events}, we implement these relations using 19 specific event subscription types to guarantee eventual consistency across the network.

\begin{table}[h!]
\centering
\caption{DDD structural metrics of the Quizzes system.}
\label{tab:ddd-structural-metrics}
\begin{tabular}{|l|c|}
\hline
\textbf{Metric} & \textbf{Value} \\ \hline
Number of aggregates & 8 \\ \hline
Number of aggregate services & 8 \\ \hline
Number of functionalities & 45 \\ \hline
Total enforced invariants & 20 \\ \hline
Number of upstream-downstream relations & 10 \\ \hline
Total subscription event types & 19 \\ \hline
\end{tabular}
\end{table}

\begin{table}[h!]
\centering
\caption{Upstream-downstream relations and event subscription counts in the Quizzes system.}
\label{tab:upstream-downstream-events}
\resizebox{\columnwidth}{!}{
\begin{tabular}{|l|l|c|}
\hline
\textbf{Upstream Aggregate} & \textbf{Downstream Aggregate} & \textbf{Event Types} \\ \hline
Topic & Question & 2 \\ \hline
User & CourseExecution & 1 \\ \hline
CourseExecution & Quiz & 1 \\ \hline
Question & Quiz & 2 \\ \hline
CourseExecution & Answer & 4 \\ \hline
Quiz & Answer & 1 \\ \hline
CourseExecution & Tournament & 4 \\ \hline
Topic & Tournament & 2 \\ \hline
Quiz & Tournament & 1 \\ \hline
Answer & Tournament & 1 \\ \hline
\textbf{Total} & & \textbf{19} \\ \hline
\end{tabular}%
}
\end{table}

The successful implementation of this diverse, logic-heavy system directly addresses RQ1. By utilizing the simulator's decoupled messaging, dynamic service discovery, and strict database-per-service isolation, we successfully support the full spectrum of DDD interactions. The \textit{Microservice Simulator} proves to be a highly expressive and complete environment, demonstrating that developers rigorously model and execute realistic aggregates across complex network topologies without compromising their underlying domain logic.

\subsection{Simulating Faults and Complex Interleavings}
\label{subsec:evaluation_faults}

The \textit{Microservice Simulator} promotes a shift-left validation strategy for microservices design.

To address RQ2 (experimenting with diverse transactional models), the simulator provides mechanisms to deterministically test how distributed coordination protocols like Sagas and Transactional Causal Consistency (TCC) behave under adverse conditions and concurrency. 

A primary challenge in validating concurrent systems is the inability to consistently replicate complex race conditions. To overcome this, the simulator allows developers to programmatically control workflow execution using the \texttt{executeUntilStep()} method in testing environments. By manually pausing one functionality mid-execution and advancing another concurrent functionality, researchers can enforce highly complex, deterministic interleavings of concurrent transactions that would be extremely difficult to isolate in a real network environment.

Furthermore, the simulator facilitates validation strategy for network boundaries. During local unit tests (e.g., utilizing Groovy and Spock), the \texttt{LocalCommandGateway} can be configured to forcefully serialize and deserialize commands and responses in-memory. This ensures that any serialization errors (such as missing default constructors or incompatible data types) are caught immediately during local development, guaranteeing that the \textit{Application Layer} payloads are strictly network-ready before deploying to distributed \textit{Stream} or \textit{gRPC} topologies.

The testing environment also enables developers to empirically measure the performance benefits of different service invocation models. To this end, the simulator was utilized to evaluate multiple asynchronous workflows. In one such test, the \texttt{CreateTournamentAsyncTest}, parallelizing cross-aggregate reads via \texttt{sendAsync()} and aggregating the \texttt{CompletableFuture} responses yielded a significant reduction in median execution time compared to strictly sequential, synchronous execution. Collectively, these asynchronous tests demonstrate the simulator's capacity to evaluate and optimize latency-sensitive business operations.

Furthermore, the \textit{Impairment Module} empowers researchers to inject business faults and execution delays dynamically. Because it integrates directly into the \textit{Microservice Simulator}'s \texttt{ExecutionPlan} orchestrator, it intercepts individual workflow steps to enforce programmed behaviors, e.g., injecting millisecond-precise thread sleeps to simulate execution delays, or forcing a \texttt{SimulatorException} to mimic a microservices failure. This capability is extensively utilized within the local Groovy testing environment, where specific failure behaviors are loaded from scripts to validate recovery mechanisms. To validate the system under load, we also executed a suite of JMeter test plans (e.g., \texttt{concurrentAddParticipant.jmx}) designed to load test the Quizzes system with conflicting requests. Crucially, the simulator distinguishes between business failures and transient infrastructure issues. While business faults (such as an injected \texttt{SimulatorException}) bypass the retry logic to immediately trigger domain-level aborts or compensations, transient network failures and serialization conflicts are transparently mitigated by the \texttt{CommandGateway}'s built-in circuit breaker (configured via \texttt{Resilience4j}). This circuit breaker intercepts recoverable errors and applies an exponential backoff retry strategy without polluting the application's domain logic.

These complex interleavings and high-load simulations allowed us to empirically contrast the resilience and developer overhead of the two supported transactional models:
\begin{itemize}
    \item \textbf{Sagas}: When coordinating with Sagas, the lack of strict isolation means concurrent operations can easily lead to anomalies such as lost updates. For example, when two concurrent requests attempt to add a participant to a tournament, the Saga model requires the developer to explicitly implement semantic locks (via \texttt{Saga State}) to flag the tournament as temporarily locked, preventing the second request from corrupting the intermediate state.
    \item \textbf{TCC}: In contrast, Transactional Causal Consistency relies on snapshot isolation and optimistic concurrency control. The same concurrent requests execute without explicit application-level locking. Instead, upon commit, the \texttt{Causal Unit Of Work Service} detects the concurrent version conflict. It automatically resolves the diverging states by applying the \texttt{mergeFields()} strategy implemented within the \texttt{Causal Tournament} aggregate, completely abstracting the anomaly resolution from the high-level business workflow.
\end{itemize}

By successfully executing these tests and resolving the induced conflicts, the simulator proves it can effectively facilitate rigorous experimentation across fundamentally different transactional models.

\subsection{Benchmarking Deployment Topologies}
\label{subsec:evaluation_benchmarks}

To directly address RQ3 (supporting diverse communication models while remaining decoupled from business logic), the simulator enables developers to benchmark the exact same application code across fundamentally different infrastructural environments. 

We automated a testing pipeline using a highly concurrent JMeter scenario to stress-test the \textit{Quizzes} system across a comprehensive spectrum of network topologies. To ensure statistical significance and account for runtime anomalies, each deployment topology was benchmarked across five independent execution runs. The primary objective was not merely to measure throughput, but to empirically demonstrate the performance variance introduced by different communication protocols when managing the overhead of distributed transactional consensus. Both supported transactional models, Sagas and Transactional Causal Consistency (TCC), were evaluated across these deployments. Specifically, the \texttt{concurrentAddParticipant.jmx} test scenario simulates a high-contention environment by bombarding the system with simultaneous requests to add different users to the exact same \texttt{Tournament} aggregate. The objective of this specific test is to aggressively and deterministically trigger concurrency conflicts. By intentionally forcing these collisions, we can empirically evaluate the overhead of each transactional model under stress, contrasting the latency of Sagas queuing requests via semantic locks against the computational cost of TCC repeatedly fetching causal snapshots and executing complex state merges.

The tested topologies ranged from simple local executions to fully decentralized microservice clusters. We focused on five core topologies to highlight the fundamental trade-offs between in-memory execution, message brokers, and remote procedure calls:
\begin{enumerate}
    \item \textbf{Centralized Local}: The baseline in-memory execution within a single JVM.
    \item \textbf{Centralized Stream}: Shared application database, but interactions routed asynchronously via a \texttt{RabbitMQ} message broker.
    \item \textbf{Centralized gRPC}: Shared application database, with interactions routed via point-to-point Remote Procedure Calls (RPC).
    \item \textbf{Distributed Stream}: Fully isolated microservices and databases communicating entirely via \texttt{RabbitMQ}.
    \item \textbf{Distributed gRPC}: Fully isolated microservices utilizing point-to-point gRPC for commands and RabbitMQ for domain events.
\end{enumerate}

To ensure robustness during these high-concurrency benchmarks, the \texttt{Command Gateway} was configured with a dedicated circuit breaker (via the Resilience4j \texttt{Retry Registry}). This mechanism transparently intercepted transient network timeouts and optimistic concurrency conflicts (which are heavily expected under concurrent load), applying an exponential backoff retry strategy. The configured maximum number of retry attempts directly influences the observed latency distributions but guarantees that temporary infrastructure bottlenecks do not manifest as hard application-level failures.

Table~\ref{tab:benchmark_results} details the empirical results of these core configurations, contrasting median latency, 95th percentile (p95) tail latency, and transaction success rate metrics across the evaluated transactional models.

\begin{table*}[t!]
\centering
\caption{Benchmark results comparing latency and success rate across deployment topologies executing a highly concurrent workflow.}
\label{tab:benchmark_results}
\resizebox{0.9\textwidth}{!}{%
\begin{tabular}{|l|l|c|c|c|c|c|c|}
\hline
\multirow{2}{*}{\textbf{Topology}} & \multirow{2}{*}{\textbf{Versioning}} & \multicolumn{3}{c|}{\textbf{Sagas}} & \multicolumn{3}{c|}{\textbf{TCC}} \\ \cline{3-8}
 & & \textbf{Med (ms)} & \textbf{p95 (ms)} & \textbf{Succ (\%)} & \textbf{Med (ms)} & \textbf{p95 (ms)} & \textbf{Succ (\%)} \\ \hline
Centralized Local  & Centralized   & 48  & 494  & 100 & 68  & 664  & 92.3 \\ \hline
Centralized Stream & Centralized   & 103 & 870  & 100 & 192 & 1453 & 96.1 \\ \cline{2-8}
                   & Decentralized & 66  & 497  & 100 & --- & ---  & ---  \\ \hline
Centralized gRPC   & Centralized   & 91  & 1156 & 100 & 181 & 1192 & 96.1 \\ \cline{2-8}
                   & Decentralized & 68  & 845  & 100 & --- & ---  & ---  \\ \hline
Distributed Stream & Centralized   & 153 & 1423 & 100 & 179 & 2027 & 96.1 \\ \cline{2-8}
                   & Decentralized & 100 & 1030 & 100 & --- & ---  & ---  \\ \hline
Distributed gRPC   & Centralized   & 133 & 2161 & 100 & 210 & 2842 & 96.1 \\ \cline{2-8}
                   & Decentralized & 99  & 1167 & 100 & --- & ---  & ---  \\ \hline
\end{tabular}%
}
\end{table*}

As the data shows, the benchmarking pipeline successfully quantifies the infrastructural overhead introduced by distributed communication protocols. Table~\ref{tab:benchmark_results} highlights several key architectural trade-offs between the Sagas and TCC models under high concurrency.

First, the purely in-memory \textit{Centralized Local} topology achieved the lowest median latency for both Sagas (48 ms) and TCC (68 ms). Transitioning to networked deployments (\textit{Centralized Stream/gRPC}) naturally increased the median latency to approximately 90--100 ms for Sagas and 180--190 ms for TCC, due to network serialization and remote command dispatching. Finally, the fully decoupled microservice topologies (\textit{Distributed Stream/gRPC}) exhibited the highest latencies, peaking at 153 ms for Sagas and 210 ms for TCC, reflecting the compounded overhead of executing multiple remote service calls to coordinate the transaction steps across isolated databases.

Second, the analysis reveals a distinct performance gap between the two transactional models. Across all deployments, Sagas maintained consistently lower median and tail latencies compared to TCC. This variance is largely attributable to TCC's optimistic concurrency control, which requires retrieving causal snapshots, performing complex state merging upon conflict detection, and interacting heavily with the centralized versioning service. Furthermore, while Sagas maintained a near-perfect success rate by relying on semantic locks to queue conflicting requests, TCC experienced a slight drop in success rate (to approximately 96.1\% remotely and 92.3\% locally), as extreme concurrency leads to persistent merge conflicts and eventual retry exhaustion.

Beyond the core topologies, the simulator also allows testing specific infrastructural optimizations, such as the \textit{Decentralized} (Snowflake) versioning profile. As shown in the table, this purely infrastructural swap yielded substantial performance gains for the Sagas model. By eliminating the centralized versioning database bottleneck, median latency in the \textit{Centralized Stream} deployment dropped by roughly 36\% (from 103 ms to 66 ms). 

Furthermore, analyzing the tail latency, specifically the 95th percentile (p95), exposes the network volatility inherent to microservice orchestration. For instance, while the median latency of the \textit{Distributed gRPC} topology for Sagas was a manageable 133 ms, its p95 spiked to 2161 ms. This severe degradation at the tail end highlights the cascading delays caused by synchronous remote procedure calls waiting on thread pool availability and network jitter during peak concurrency. However, applying the \textit{Decentralized} Snowflake ID generator to the \textit{Distributed gRPC} topology nearly halved its p95 tail latency (from 2161 ms down to 1167 ms) by eliminating the synchronous database lock contention at the centralized versioning service. Crucially, as previously noted, this decentralized optimization is fundamentally incompatible with TCC, which is constrained to centralized versioning and thus cannot benefit from this latency reduction.

Crucially, transitioning between these drastically different infrastructural benchmarks, and achieving these optimizations, required zero modifications to the underlying \textit{Quizzes} domain logic or the transactional workflows. By simply altering the application's configuration profiles, the simulator dynamically swapped the underlying \texttt{Command Gateway} and \texttt{Version Service} implementations. This unequivocally demonstrates the simulator's success in cleanly decoupling complex business logic from the underlying communication and deployment mechanics, providing researchers with a seamless environment to evaluate architectural trade-offs early in the development lifecycle.

\subsection{Developer Effort and Extensibility}
\label{subsec:evaluation_effort}

To address RQ4, we evaluate the effort required to utilize and extend the simulator's architecture. Because the architectural modules (detailed in Section~\ref{sec:microservice-simulator}) heavily utilize structural design patterns (e.g., Decorators and Template Methods) to isolate infrastructural concerns, the cognitive and coding burden placed on developers is significantly minimized. We analyze this effort across two distinct developer personas.

\textbf{Application Developer Effort:}
For an application developer, the simulator abstracts the complexities of distributed infrastructure. The developer workflow is strictly focused on domain modeling, defining events, and writing the business logic. Tables~\ref{tab:developer_effort_domain} and \ref{tab:developer_effort_functionality} detail the specific steps required to implement an aggregate and a functionality using the simulator, illustrating the separation between the domain logic provided by the developer and the infrastructural coordination handled by the simulator. While the majority of the effort is focused on the microservice being developed, supporting centralized deployment topologies (such as \textit{Centralized Stream} or \textit{Centralized gRPC}) introduces a minor configuration step: the developer must append the new aggregate's specific network bindings (e.g., stream channels and event broker bindings) into a shared, unified \texttt{application.yaml} file to ensure the centralized execution environment can correctly route messages between the aggregates.

\begin{table*}[t!]
\centering
\caption{Application Developer Effort within the Application Domain Module: Implementing a Single Aggregate (\texttt{Tournament}).}
\label{tab:developer_effort_domain}
\resizebox{0.9\textwidth}{!}{%
\begin{tabular}{|p{3.6cm}|p{7.0cm}|p{5.4cm}|}
\hline
    \textbf{Development Task} & \textbf{Implementation Details \& Example} & \textbf{Rationale (Why)} \\ \hline

\multicolumn{3}{|c|}{\textbf{1. Service Scaffolding}} \\ \hline
Define Spring Boot Application & Create the microservice entry point, e.g., \href{https://github.com/socialsoftware/microservices-simulator/blob/v3.2.0/applications/quizzes/src/main/java/pt/ulisboa/tecnico/socialsoftware/quizzes/microservices/tournament/TournamentServiceApplication.java}{\texttt{TournamentServiceApplication.java}} with \texttt{ @SpringBootApplication}. & Establishes the bounded context runtime and independent deployability. \\ \hline
Define DTOs and Repositories & Create data transfer objects and Spring Data JPA interfaces for data access, e.g., \href{https://github.com/socialsoftware/microservices-simulator/blob/v3.2.0/applications/quizzes/src/main/java/pt/ulisboa/tecnico/socialsoftware/quizzes/microservices/tournament/aggregate/TournamentDto.java}{\texttt{TournamentDto.java}}, \href{https://github.com/socialsoftware/microservices-simulator/blob/v3.2.0/applications/quizzes/src/main/java/pt/ulisboa/tecnico/socialsoftware/quizzes/microservices/tournament/aggregate/TournamentRepository.java}{\texttt{TournamentRepository.java}}. & Separates persistence/API contracts from domain behavior and supports query/update paths. \\ \hline
Define Aggregate Services & Define the microservice API to register changes, e.g., \href{https://github.com/socialsoftware/microservices-simulator/blob/v3.2.0/applications/quizzes/src/main/java/pt/ulisboa/tecnico/socialsoftware/quizzes/microservices/tournament/service/TournamentService.java}{\texttt{updateUserName(...)}}. & Provides stable operation-level contracts used by commands and controllers. \\ \hline
Define Web Controllers & Expose REST API endpoints to external clients, e.g., \href{https://github.com/socialsoftware/microservices-simulator/blob/v3.2.0/applications/quizzes/src/main/java/pt/ulisboa/tecnico/socialsoftware/quizzes/microservices/tournament/coordination/webapi/TournamentController.java}{\texttt{TournamentController.java}}. & Enables external access while preserving application/domain layering. \\ \hline

\multicolumn{3}{|c|}{\textbf{2. Aggregate Consistency}} \\ \hline
Define Aggregate & Define the JPA root entity, e.g., \href{https://github.com/socialsoftware/microservices-simulator/blob/v3.2.0/applications/quizzes/src/main/java/pt/ulisboa/tecnico/socialsoftware/quizzes/microservices/tournament/aggregate/Tournament.java}{\texttt{Tournament.java}}, and associated value objects, e.g., \href{https://github.com/socialsoftware/microservices-simulator/blob/v3.2.0/applications/quizzes/src/main/java/pt/ulisboa/tecnico/socialsoftware/quizzes/microservices/tournament/aggregate/TournamentCreator.java}{\texttt{TournamentCreator}}. & Defines the transactional consistency boundary where invariants are enforced. \\ \hline
Specify Invariants & Override the \href{https://github.com/socialsoftware/microservices-simulator/blob/v3.2.0/applications/quizzes/src/main/java/pt/ulisboa/tecnico/socialsoftware/quizzes/microservices/tournament/aggregate/Tournament.java}{\texttt{verifyInvariants()}} method, e.g., asserting tournament start date is before end date. & Defines the aggregate consistency rules. \\ \hline
Define Transactional Aggregates & Extend aggregate for specific models, e.g., \href{https://github.com/socialsoftware/microservices-simulator/blob/v3.2.0/applications/quizzes/src/main/java/pt/ulisboa/tecnico/socialsoftware/quizzes/microservices/tournament/aggregate/sagas/SagaTournament.java}{\texttt{SagaTournament.java}} (locks) and \href{https://github.com/socialsoftware/microservices-simulator/blob/v3.2.0/applications/quizzes/src/main/java/pt/ulisboa/tecnico/socialsoftware/quizzes/microservices/tournament/aggregate/causal/CausalTournament.java}{\texttt{CausalTournament.java}} (merging). & Adapts the same domain to model-specific consistency semantics without duplicating business logic. \\ \hline
Define Commands & Define remote commands for aggregate services, e.g., \href{https://github.com/socialsoftware/microservices-simulator/blob/v3.2.0/applications/quizzes/src/main/java/pt/ulisboa/tecnico/socialsoftware/quizzes/commands/tournament/AddParticipantCommand.java}{\texttt{AddParticipantCommand.java}}. & Formalizes inter-service invocation contracts independent of transport protocol. \\ \hline
Create CommandHandler & Receive remote commands and map to services, e.g., \href{https://github.com/socialsoftware/microservices-simulator/blob/v3.2.0/applications/quizzes/src/main/java/pt/ulisboa/tecnico/socialsoftware/quizzes/microservices/tournament/messaging/TournamentCommandHandler.java}{\texttt{TournamentCommandHandler.java}}. & Centralizes command routing and isolates transport concerns from domain services. \\ \hline

\multicolumn{3}{|c|}{\textbf{3. Event-Based Integration}} \\ \hline
Define Events & Define the events published/subscribed, e.g., \href{https://github.com/socialsoftware/microservices-simulator/blob/v3.2.0/applications/quizzes/src/main/java/pt/ulisboa/tecnico/socialsoftware/quizzes/events/UpdateStudentNameEvent.java}{\texttt{UpdateStudentNameEvent.java}}. & Makes upstream changes observable by downstream aggregates for eventual consistency. \\ \hline
Subscribe Events & Override the \href{https://github.com/socialsoftware/microservices-simulator/blob/v3.2.0/applications/quizzes/src/main/java/pt/ulisboa/tecnico/socialsoftware/quizzes/microservices/tournament/aggregate/Tournament.java}{\texttt{getEventSubscriptions()}} method, adding concrete subscriptions. & Declares upstream-to-downstream dependencies explicitly at the domain level. \\ \hline
Define Event Subscriptions & Define subscription conditions, e.g., in \href{https://github.com/socialsoftware/microservices-simulator/blob/v3.2.0/applications/quizzes/src/main/java/pt/ulisboa/tecnico/socialsoftware/quizzes/microservices/tournament/notification/subscribe/TournamentSubscribesUpdateStudentName.java}{\texttt{TournamentSubscribesUpdateStudentName.java}} a tournament subscribes to creator/participant name updates. & Filters only relevant upstream events, avoiding unnecessary or inconsistent updates. \\ \hline
Define Event Handlers & Delegate handling to processing functionalities, e.g., \href{https://github.com/socialsoftware/microservices-simulator/blob/v3.2.0/applications/quizzes/src/main/java/pt/ulisboa/tecnico/socialsoftware/quizzes/microservices/tournament/notification/handling/handlers/UpdateStudentNameEventHandler.java}{\texttt{UpdateStudentNameEventHandler.java}}. & Associates a functionality to handle the event. \\ \hline
Define Event Handling & Define polling logic for the event table, e.g., \href{https://github.com/socialsoftware/microservices-simulator/blob/v3.2.0/applications/quizzes/src/main/java/pt/ulisboa/tecnico/socialsoftware/quizzes/microservices/tournament/notification/handling/TournamentEventHandling.java}{\texttt{TournamentEventHandling.java}}. & Drives periodic event detection for event handling. \\ \hline
Define Event Subscriber Service & Define the function definition name in the yaml file, e.g., in \href{https://github.com/socialsoftware/microservices-simulator/blob/v3.2.0/applications/quizzes/src/main/resources/application-tournament-service.yaml}{\texttt{application-tournament-service.yaml}}. & Bridges broker transport to local event persistence/processing. \\ \hline

\multicolumn{3}{|c|}{\textbf{4. Deployment Wiring}} \\ \hline
Configure Network Bindings & Set \texttt{stream} channels or \texttt{grpc} ports in the aggregate yaml file, e.g., in \href{https://github.com/socialsoftware/microservices-simulator/blob/v3.2.0/applications/quizzes/src/main/resources/application-tournament-service.yaml}{\texttt{application-tournament-service.yaml}} and append them to the unified \href{https://github.com/socialsoftware/microservices-simulator/blob/v3.2.0/applications/quizzes/src/main/resources/application.yaml}{\texttt{application.yaml}}. & Activates deployment topologies (both distributed and centralized) without changing business code. \\ \hline
Configure API Gateway Routes & Define route mappings in the microservice yaml to route HTTP requests, e.g., in \href{https://github.com/socialsoftware/microservices-simulator/blob/v3.2.0/applications/quizzes/src/main/resources/application-tournament-service.yaml}{\texttt{application-tournament-service.yaml}}. & Decouples external API paths from internal service locations. \\ \hline
\end{tabular}
}
\end{table*}

\begin{table*}[t!]
\centering
\caption{Application Developer Effort within the Application Functionality Module: Implementing a Single Functionality (\texttt{AddParticipant}).}
\label{tab:developer_effort_functionality}
\resizebox{0.9\textwidth}{!}{%
\begin{tabular}{|p{3.6cm}|p{7.3cm}|p{5.1cm}|}
\hline
    \textbf{Development Task} & \textbf{Implementation Details \& Example} & \textbf{Rationale (Why)} \\ \hline
Define Functionality & Extend \href{https://github.com/socialsoftware/microservices-simulator/blob/v3.2.0/simulator/src/main/java/pt/ulisboa/tecnico/socialsoftware/ms/coordination/WorkflowFunctionality.java}{\texttt{WorkflowFunctionality}} to coordinate a specific use-case, e.g., \href{https://github.com/socialsoftware/microservices-simulator/blob/v3.2.0/applications/quizzes/src/main/java/pt/ulisboa/tecnico/socialsoftware/quizzes/microservices/tournament/coordination/sagas/AddParticipantFunctionalitySagas.java}{\texttt{AddParticipantFunctionalitySagas.java}}. & Encapsulates one business use case as a reusable coordination unit. \\ \hline
Workflow Orchestration & Map execution \href{https://github.com/socialsoftware/microservices-simulator/blob/v3.2.0/simulator/src/main/java/pt/ulisboa/tecnico/socialsoftware/ms/coordination/Step.java}{\texttt{Step}}s, dependencies, and transaction triggers within \href{https://github.com/socialsoftware/microservices-simulator/blob/v3.2.0/applications/quizzes/src/main/java/pt/ulisboa/tecnico/socialsoftware/quizzes/microservices/tournament/coordination/sagas/AddParticipantFunctionalitySagas.java}{\texttt{buildWorkflow()}}, e.g., defining \texttt{getUserStep} and \texttt{addParticipantStep} dependencies. & Makes ordering, dependency, and rollback/compensation boundaries explicit. \\ \hline
Command Dispatching & Instantiate remote \href{https://github.com/socialsoftware/microservices-simulator/blob/v3.2.0/simulator/src/main/java/pt/ulisboa/tecnico/socialsoftware/ms/messaging/Command.java}{\texttt{Command}}s and dispatch via the abstract \href{https://github.com/socialsoftware/microservices-simulator/blob/v3.2.0/simulator/src/main/java/pt/ulisboa/tecnico/socialsoftware/ms/messaging/CommandGateway.java}{\texttt{CommandGateway}}, e.g., sending \href{https://github.com/socialsoftware/microservices-simulator/blob/v3.2.0/applications/quizzes/src/main/java/pt/ulisboa/tecnico/socialsoftware/quizzes/commands/tournament/AddParticipantCommand.java}{\texttt{AddParticipantCommand}} wrapped in a \href{https://github.com/socialsoftware/microservices-simulator/blob/v3.2.0/simulator/src/main/java/pt/ulisboa/tecnico/socialsoftware/ms/transaction/sagas/messaging/SagaCommand.java}{\texttt{SagaCommand}} with semantic locks. & Executes distributed steps through transport-agnostic contracts while preserving domain isolation. \\ \hline
\end{tabular}
}
\end{table*}

Table~\ref{tab:developer_effort_domain} decomposes effort into four layers: service scaffolding, aggregate consistency, event-based integration, and deployment wiring. Most effort is concentrated in the domain and integration artifacts, while infrastructure switching remains primarily configuration-driven.

Table~\ref{tab:developer_effort_functionality} shows that application-level effort is mostly orchestration design rather than infrastructure coding. The developer specifies dependency structure and command intent, while execution mechanics, communication, and transaction wrapping are delegated to the simulator.

\textbf{Simulator Library Developer Effort:}
Conversely, evaluating the simulator's extensibility involves analyzing the effort required by a library developer to introduce entirely new infrastructural models. Because the \textit{Business Layer} relies on strict interfaces, extensions require implementing localized plugins rather than modifying existing applications.

\begin{itemize}
    \item \textbf{Adding Communication Models:} As established in Section~\ref{subsec:messaging_module}, injecting a new protocol (e.g., REST) fundamentally requires implementing a new \texttt{Command Gateway} to dispatch commands and a corresponding receiver service. However, due to the specific quirks of different communication models, library developers may need to introduce supplementary components to orchestrate the send and receive lifecycles. For instance, implementing the \texttt{stream} profile required additional infrastructure, such as a \texttt{Command Response Aggregator} and dedicated response listeners, to seamlessly simulate synchronous request-reply semantics over an inherently asynchronous, decoupled message broker. Crucially, regardless of this underlying transport complexity, the library developer does not need to write cross-cutting logic. The existing \texttt{Command Handler Decorator} intercepts the execution loop to wrap the transactional logic, ensuring that the transport mechanics remain completely isolated from the application's \texttt{handleDomainCommand()}.
    \item \textbf{Adding Transactional Models:} As established in Section~\ref{subsec:transaction_module}, introducing a transactional model requires extending the \textit{Transaction} and \textit{Coordination} modules. A library developer must implement several protocol-specific counterparts: a new \texttt{Unit Of Work Service} and \texttt{Unit Of Work} to manage transactional state; a handler extending \texttt{Transaction Command Handler} to intercept incoming commands and execute transaction-specific logic (e.g., register semantic locks in \textit{Sagas}) before and after domain execution; protocol-specific \texttt{Workflow} and \texttt{Step} implementations; and a new wrapper \texttt{Command} and marker \texttt{Aggregate} interface to enforce application-level compatibility. This strict separation of concerns ensures that extending the simulator's capabilities with entirely new transactional theories does not demand sweeping refactors of the deployed microservices.
\end{itemize}

\subsection{Threats to Validity}
\label{subsec:evaluation_threats}

To rigorously assess the limitations of our evaluation and the extent to which the research questions (RQs) were validated, we analyze the threats to validity across three dimensions: construct, internal, and external validity.

\subsubsection{Construct Validity:} 
Construct validity concerns whether the experimental setup accurately measures the concepts defined in our RQs. To validate RQ1 (facilitating DDD) and RQ4 (minimizing developer effort), we analyzed the implementation of the \textit{Quizzes} system and quantified the required developer actions. A primary threat here is subjective bias, as the application was extended by the simulator's authors. However, this subjective bias is mitigated by the specific design focus of the \textit{Quizzes} application. Standard microservice benchmarking suites, such as TeaStore~\cite{teastore2018} and DeathStarBench~\cite{deathstarbench2019}, feature realistic applications but are primarily designed to evaluate hardware, operating system, and network-level performance, such as resource scaling and tail-latency anomalies. In contrast, the \textit{Quizzes} system was explicitly designed to test Domain-Driven Design (DDD) semantics. It exercises complex inter-aggregate dependencies, strict business invariants, and the specific conflict resolution mechanisms required by distributed transactional models like Sagas and TCC. This ensures that the simulator was validated against a scenario where domain-heavy consistency challenges are the primary focus.

Furthermore, to validate RQ2 (transactional models under adverse conditions), we utilized the \texttt{executeUntilStep()} method and the \textit{Impairment Module} to inject deterministic delays and business faults. While this effectively triggers optimistic concurrency conflicts and recovery mechanisms, programmatically paused threads and simulated exceptions may not perfectly encapsulate the unpredictable and chaotic nature of arbitrary hardware crashes or network partitions found in real-world distributed systems~\cite{bailis2014network, gunawi2016does}. However, our messaging evaluation is intentionally focused on business-level resilience. For instance, an orchestrator crash during a Saga's commit loop can leave remote aggregates permanently holding semantic locks, while a failure during TCC's Two-Phase Commit can leave participants in an indefinitely blocked prepared state. These stronger guarantees require additional protocol and persistence mechanisms that are outside the current scope of this work. Nevertheless, the current implementation accomplishes the core goals of the research questions. By explicitly isolating these transport-level concerns, the simulator provides a sufficiently robust environment to directly evaluate and compare the conflict resolution capabilities of different transactional models (RQ2), assess the architectural overhead across diverse communication topologies (RQ3), and quantify the required developer effort (RQ4) at the business logic level.

A potential threat to construct validity lies in the architectural mapping of Domain-Driven Design concepts to the simulated deployment. In traditional DDD, a \textit{Bounded Context} often encapsulates multiple aggregates managed by a single team~\cite{evans2004domain}. However, the \textit{Quizzes} evaluation strictly enforces a one-to-one mapping between a single aggregate and an independent microservice. While this architectural simplification means the evaluation does not model \textit{Bounded Contexts} containing multiple local aggregates, it is a deliberate methodological choice. The aggregate, by definition, is the unit of atomicity in the interaction between microservices, while the bounded context is the unit of development between different teams. Therefore, by associating every aggregate with a distinct microservice, the simulator focuses on its goal, the design and analysis of the behavior of business logic in the context of distributed complexity. If we want to experiment with atomic transactional behavior across multiple aggregates, we can wrap them within the same microservice, allowing commands to interact with more than one aggregate simultaneously.

\subsubsection{Internal Validity:} 
Internal validity concerns whether hidden or uncontrolled factors might have skewed our empirical results. To address RQ3 (supporting diverse topologies), we shown how easy it is to compare latency, throughput, and success rates of the system under high load using JMeter for different technologies. A significant internal threat stems from resource contention during these benchmarks. Because the simulated microservices, databases, and message brokers were executed within containerized local environments, shared resources such as CPU limits, memory constraints, network stack limits, and JVM Garbage Collection (GC) pauses could inadvertently introduce latency spikes or bottleneck the RabbitMQ broker. While benchmarking across five independent runs mitigates some runtime variance, these shared-host constraints dictate that the absolute latency values reported are less indicative than the relative performance overhead observed between the different topologies. However, because the primary objective of the evaluation's benchmark is the comparative analysis of architectural trade-offs rather than establishing latency baselines for each deployment, this shared-host contention represents an acceptable limitation that does not invalidate the core comparative findings.

\subsubsection{External Validity:} 
External validity assesses the degree to which our findings can be generalized to the broader software industry. To validate the simulator's capabilities across all RQs, we utilized the \textit{Quizzes} application, a realistic, business-logic-rich domain model heavily focused on complex orchestration and workflows. The primary external threat is that microservice architectures span a vast array of domain typologies. It is uncertain if the simulator's decoupled abstractions, aggregate boundaries, and transactional coordination mechanisms (Sagas and TCC) would remain equally effective for domains with drastically different data access patterns or varying degrees of business logic complexity. Additionally, because security and reliability assumptions vary widely across domains, extensions such as transport security and network-level failure modeling are left for future work before claiming broader industrial generalization. Crucially, while these infrastructural hardening aspects are necessary for full production readiness, their absence does not detract from the simulator's core goal: enabling the rigorous, shift-left architectural validation of business logic and distributed data consistency.

\section{Conclusion}
\label{sec:conclusion}

The proposed \textit{Microservice Simulator} presents a robust, extensible architecture for managing the inherent complexities of developing business-logic-rich distributed systems. By rooting the design in strict Domain-Driven Design (DDD) principles and enabling a shift-left verification strategy, the simulator significantly reduces the risks and costs associated with late-stage data consistency discovery. It empowers developers to identify, simulate, and resolve coordination bottlenecks early in the software development lifecycle.

Our evaluation indicates that the simulator effectively handles intricate aggregate dependencies in the evaluated scenarios, supporting invariant preservation and conflict resolution. Unlike existing performance-centric simulators and production-focused execution frameworks, this simulator successfully supports rigorous architectural validation by executing the actual domain logic across fundamentally different transactional paradigms, efficiently managing both Sagas and Transactional Causal Consistency (TCC). Furthermore, the programmatic testing capabilities allow researchers to simulate complex, deterministic interleavings of distributed transactions, while the integration of the \textit{Impairment Module} enables dynamic fault injection and delays, providing empirical evidence of resilience for these models under the adverse conditions evaluated in this study.

Crucially, this work bridges the gap between local simulation and real-world execution. By strictly decoupling the pure domain logic from the underlying infrastructure, the simulator allows developers to evaluate their systems across diverse communication models. As demonstrated through empirical benchmarking, the exact same business workflows can be seamlessly transitioned from a local, in-memory testing sandbox to fully decentralized, containerized topologies utilizing gRPC, RabbitMQ, and centralized or Snowflake-based versioning. This enables development teams to empirically measure coordination overheads, median latencies, and scalability trade-offs without rewriting any application code.

Future development should focus on expanding the simulator's operational versatility, analytical power, and developer experience. To further validate the simulator's adaptability, subsequent research could involve modeling additional complex applications across diverse domains. Furthermore, future studies should investigate the orchestration of multi-aggregate \textit{Bounded Contexts}. While the current evaluation enforces a strict one-to-one mapping between aggregates and microservices, extending the simulator to manage atomic transactional behavior across multiple aggregates within the same microservice will provide deeper insights into traditional DDD modeling. Operationally, the simulator's library can be extended to encompass new transactional models and alternative messaging protocols, providing researchers with a broader toolkit for architectural evaluation.

With the core research questions addressed, future work can explore extending the simulator's infrastructural realism by strengthening the underlying transport guarantees. As an extension to the current messaging module, future iterations could introduce stable command identifiers to support exactly-once idempotency. Coupled with these messaging upgrades, the addition of a distributed transaction recovery manager presents an avenue for future research. Such a component would asynchronously detect stalled orchestrators and execute timeout-based rollbacks or commit-resumes, further enhancing resilience against semantic lock starvation in Sagas and blocked participant states in TCC's Two-Phase Commit. Finally, to extend the simulator's applicability to strict operational environments, future work could explore the integration of transport-security and service-authentication mechanisms.

Building upon the validation achieved in this work, the simulator's testing and analytical capabilities present avenues for future enhancement. While the current \textit{Impairment Module} successfully evaluates business-level fault tolerance, future extensions could model severe network-level degradations, such as arbitrary broker partitions and transient unavailability waves, to stress-test system resilience under extreme infrastructural chaos. Also, the existing observability infrastructure provides a strong foundation that future iterations could augment with automated evaluation tools, designed to proactively flag performance anomalies, excessive latency, and coordination bottlenecks. Finally, future work could further optimize this experience by integrating code generation techniques. By automatically producing the boilerplate for domain entities, functionalities, and infrastructural bindings, developers could focus more on higher-level application structure and strategic system design.

Ultimately, the \textit{Microservice Simulator} provides a highly realistic, developer-friendly environment to investigate coordination mechanisms, evaluate the viability of architectural boundaries, and make well-informed design decisions before committing to expensive production implementations.

To ensure the reproducibility of our evaluation and benchmarks, the exact version of the \textit{Microservice Simulator} and the \textit{Quizzes} case study used in this paper has been permanently archived. The source code is available via the GitHub release tag \texttt{v3.2.0} at: \url{https://github.com/socialsoftware/microservices-simulator/releases/tag/v3.2.0}.

\section*{Acknowledgment}

\sloppy
\emergencystretch=1em

\noindent
\small This research builds upon the foundational codebase of the \textit{Microservice Simulator}. The core transactional coordination mechanisms were established by Pedro Pereira, who developed the Transactional Causal Consistency (TCC) model~\cite{pereira2023transactional, pereira2022transactional}, and André Martins Esgalhado, who integrated the Saga orchestration pattern~\cite{esgalhado2024simulator}. Furthermore, the empirical evaluation presented in this work was made possible by the recent infrastructural testing extensions developed by Tomás Nascimento~\cite{nascimento2024fault} and Rómulo Kaidussis~\cite{kaidussis2024performance}, who implemented the Monitoring and Impairment modules, respectively. \\
Work supported by national funds through Fundação para a Ciência e a Tecnologia, I.P. (FCT) under projects UID/50021/2025 (DOI: \href{https://doi.org/10.54499/UID/50021/2025}{https://doi.org/10.54499/UID/50021/2025} ) and UID/PRR/50021/ 2025 (DOI: \href{https://doi.org/10.54499/UID/PRR/50021/2025}{https://doi.org/10.54499/UID/PRR/50021/2025}).

\bibliographystyle{unsrt}
\bibliography{main}

\end{document}